\documentclass{llncs}
 
\usepackage{amsmath,amssymb}
\usepackage{QED}
\usepackage[all]{xy}
\usepackage{prooftree}

\usepackage{pstricks,pst-node}
 \psset{rowsep=.1cm,colsep=.1cm}
 \psset{arrowsize=3.5pt 3.8,arrowlength=0.7}
 \psset{labelsep=.1cm,nodesep=2pt}
\def\_{\kern.08em\vbox{\hrule width.35em height.6pt}\kern.08em}

\newcommand{\coqdocindent}[1]{\noindent\kern#1}

\newtheorem{defn}{Definition}

\newtheorem{lem}[defn]{Lemma}
\newtheorem{thm}[defn]{Theorem}
\newtheorem{cor}[defn]{Corollary}

\newcommand{\eqdef}{\stackrel{\Delta}{=}}
\newcommand{\sff}{F}

\author{St\'ephane Le Roux\thanks{http://perso.ens-lyon.fr/stephane.le.roux/. Now working at INRIA-Microsoft Research. I thank Pierre Lescanne for his comments on the draft of this paper.}}
 
\institute{\'Ecole normale sup\'erieure de Lyon, Universit\'e de Lyon, LIP, CNRS, INRIA, UCBL}

\title{Graphs and Path Equilibria}

\begin{document}

\maketitle

\begin{abstract}
The quest for optimal/stable paths in graphs has gained attention in a few practical or theoretical areas. To take part in this quest this chapter adopts an equilibrium-oriented approach that is abstract and general: it works with (quasi-arbitrary) arc-labelled digraphs, and it assumes very little about the structure of the sought paths and the definition of equilibrium, \textit{i.e.} optimality/stability. In this setting, this chapter presents a sufficient condition for equilibrium existence for every graph; it also presents a necessary condition for equilibrium existence for every graph. The necessary condition does not imply the sufficient condition a priori. However, the chapter pinpoints their logical difference and thus identifies what work remains to be done. Moreover, the necessary and the sufficient conditions coincide when the definition of optimality relates to a total order, which provides a full-equivalence property. These results are applied to network routing.
\end{abstract}

\section{Introduction}

This chapter provides an abstract formalism that enables generic proofs, yet accurate results, about path equilibria in graphs. For other approaches to optimisation in graphs see~\cite{Lawler01}, for instance. Beyond this, the purpose of this chapter is to provide a tool for a generalisation of sequential (tree-) games within graphs. However, these game-theoretic facets are not discussed in this chapter. In addition to the game-theoretic application, the results presented in this chapter may help solve problems of optimisation/stability of paths in graphs: a short example is presented for the problem of network routing .

\subsection{Contribution}

This chapter introduces the terminology of \emph{dalographs} which refers to finite, arc-labelled, directed graphs with non-zero outdegree, \textit{i.e.} each of whose node has an outgoing arc. An embedding of arc-labelled digraphs into dalographs shows that the non-zero-outdegree constraint may not yield a serious loss of generality. The paths that are considered in this chapter are infinite. Indeed, finite paths and infinite paths are of slightly different "types". Considering both may hinder an algebraic approach of the system. However, another embedding allows representing finite paths in a dalograph as infinite paths in another dalograph. This shows that the infiniteness constraint may not yield a serious loss of generality either. Note that the non-zero-outdegree constraint ensures existence of infinite paths, starting from any node. This uniformity facilitates an algebraic approach of the system. The paths considered in this chapter are non-self-crossing, which somehow suggests consistency. This sounds desirable in many areas, but it may be an actual restriction in some others.

In this formalism, a path induces an ultimately periodic sequence of labels (of arcs that are involved in the path). An arbitrary binary relation over ultimately periodic sequences of labels is assumed and named \emph{preference}. This induces a binary relation over paths, which is also named preference. It is defined as follows. Given two paths starting from the same node, one is preferred over the other if the sequence of labels that is induced by the former is preferred over the sequence of labels that is induced by the latter. Maximality of a given path in a graph means that no path is preferred over the given path. A \emph{strategy} is an object built over a dalograph. It amounts to every node choosing an outgoing arc. This way, a strategy induces paths starting from any given node. An \emph{equilibrium} is a strategy inducing optimal paths for any node. 

The proof of equilibrium existence is structured as follows. First, a \emph{seeking-forward} function is defined so that given a node it returns a path. Given a node, the function chooses a path that is maximal (according to the definition of the previous paragraph), and the function "follows" the path until the remaining path is not maximal (among the paths starting from the current node). In this case, the function chooses a maximal path starting from the current node and proceeds as before. All of this is done under the constraint that a path is non-self-crossing. Under some conditions, this procedure yields a path that is maximal not only at its starting node, but also at all nodes along the path. Such a path is called a \emph{hereditary maximal path}. Equipped with this lemma, the existence of an equilibrium for every dalograph is proved as follows by induction on the number of arcs in the dalograph.

\begin{itemize}
\item Compute a hereditary maximal path in the dalograph.
\item Remove the arcs of the dalograph that the path ignored while visiting adjacent nodes and get a smaller dalograph. 
\item Compute an equilibrium on this smaller dalograph and add the ignored arcs back. This yields an equilibrium for the original dalograph. 
\end{itemize}

The sufficient condition for equilibrium existence involves a notion lying between strict partial order and strict total order, namely \emph{strict weak order}, which is discussed in~\cite{Fishburn70}, for instance. This chapter requires a few preliminary results about strict weak orders. Moreover, the definition of the seeking-forward function requires the design of a recursion principle that is also used as a proof principle in this chapter. To show the usefulness of this sufficient condition, this chapter provides a few examples of non-trivial relations that meet the requirements of this condition: lexicographic extension of a strict weak order, Pareto-like order, and two limit-set-oriented orders. Then, as an application to network routing, one derives a sufficient condition for a routing policy to guarantee existence of stable routing solutions.

The proof of the necessary condition for equilibrium existence involves various closures of binary relations. Most of the closures defined here are related to properties that are part of the sufficient condition. For instance, the sufficient condition involves transitivity of some preference (binary relation), and the necessary condition involves the transitive closure of the preference. Each of those various closures is proved to preserve existence of equilibrium. More specifically, if a preference ensures existence of equilibrium for every dalograph, then the closure of the preference also ensures existence of equilibrium for every dalograph. A new closure is defined subsequently as the combination of all the above-mentioned closures. Thanks to a preliminary development introducing the notion of \emph{simple closure}, this combination closure also preserves existence of equilibrium. Since the combination closure has a few properties, this gives a non-trivial necessary condition for equilibrium existence. All the closures mentioned above are defined inductively through inference rules, which allows using rule induction as a proof principle. To show the usefulness of this necessary condition, this chapter provides an example of a non-trivial relation that does not meet the requirements of this condition.

However, not all the properties that are part of the sufficient condition can be easily turned into a simple closure preserving equilibrium existence (and hereby be part of the necessary condition). Therefore, the necessary condition and sufficient condition do not look exactly the same. Some examples show that the necessary condition is too weak. Some more work, \textit{i.e.} the design of more general simple closures, is likely to take care of those examples, and hereby provide a stronger necessary condition. However, there is also likely to be more complex examples that would require the design of more complex simple closures. As to the sufficient condition, it is still unclear whether or not it is too strong. 

In the case where the preference is a total order, the sufficient and necessary conditions coincide. This gives a necessary and sufficient condition (on a total order preference) for existence of equilibrium in every dalograph. This leads to a necessary and sufficient condition (on a total order preference) for equilibrium existence in the network routing application mentioned above. 
 
\subsection{Contents}

Section~\ref{sect:dalo-eq} defines a graph-like structure named dalograph and the notion of equilibrium in dalographs. Section~\ref{sect:br} defines a refinement of partial order that is named strict weak order in the literature. This section also connects the notion of strict weak order to other properties of binary relations. Section~\ref{sect:eq-ex} proves a sufficient condition that guarantees existence of equilibrium for all dalographs. It also gives a few examples of non-trivial relations meeting the requirements of the sufficient condition. Finally, it gives an application in network routing. Section~\ref{sect:pc} defines the notion of simple closure and proves a property about the \emph{union} of simple closures. Section~\ref{sect:p-eqex} discusses preservation of equilibrium existence by closures, and thus provides a necessary condition for existence of equilibrium for all dalographs. Section~\ref{sect:sc-nc} compares the sufficient condition and the necessary condition, and shows that they coincide in the total order case but not in general. It also describes a relation that does not meet the requirements of the necessary condition. Finally, it gives a necessary and sufficient condition, in the total order case, for equilibrium existence in the network routing application mentioned above.  

\subsection{Conventions}

Unless otherwise stated, universal quantifiers are usually omitted in front of (semi-) formal statements. For instance, a claim of the form $P(x,y)$ should be read $\forall x,y,\,P(x,y)$. 

Usually, when proving a claim of type $\tau_0\Rightarrow\dots\Rightarrow\tau_{n-1}\Rightarrow\tau_n\Rightarrow\tau$, the beginning of the proof implicitly assumes $\tau_0$,..., $\tau_{n-1}$ , and $\tau_n$, and then starts proving $\tau$. 

Given $u$ a non-empty finite sequence, $u^f$ (resp. $u^l$) represents the first (resp. last) element of $u$.

The notation $P(x)\,\eqdef\,Q(x)$ means that $P$ is defined as coinciding with $Q$.

The negation of a relation $\prec$ is written as follows.

\[\alpha\not\prec\beta\quad\eqdef\quad\neg(\alpha\prec\beta)\]

The corresponding incomparability relation is defined below.

\[\alpha\sharp\beta\quad\eqdef\quad\alpha\not\prec\beta\,\wedge\,\beta\not\prec\alpha\]

The inverse of a relation $\prec$ is defined as follows.

\[\alpha\succ\beta\quad\eqdef\quad\beta\prec\alpha\]

Restrictions of binary relations are written as follows.

 \[\alpha\prec\mid_S\beta\quad\eqdef\quad \alpha\prec \beta\,\wedge\, (\alpha,\beta)\in S^2\]

Function composition is defined as (almost) usual. For any functions $f_1$ of type $A\to B$ and $f_2$ of type $B\to C$, the composition of $f_1$ and $f_2$ is written $f_2\circ f_1$. It is of type $A\to C$ and it is defined as follows.

\[f_2\circ f_1(x)\quad\eqdef\quad f_2(f_1(x))\]

In this chapter, $f_1\circ f_2$ may be written $f_1f_2$, in the same order.

\section{Dalographs and Equilibria}\label{sect:dalo-eq}

Subsection~\ref{subsect:dalo} defines a class of arc-labelled digraphs, subsection~\ref{subsect:w-p} defines a class of paths in such arc-labelled digraphs, and subsection~\ref{subsect:eq} derives a notion of equilibrium from a notion of maximality for paths. The more general will be those graphs and paths, the more general will be the derived notion of equilibrium. 

\subsection{Dalographs}\label{subsect:dalo}

This subsection gives a name to the class of arc-labelled directed graphs each of whose node has at least one outgoing arc. Then it briefly justifies why arc labelling is "more general" than node labelling, and why the non-zero-outdegree constraint may not yield a serious loss of generality.

\begin{defn}[Dalograph]
Let $L$ be a collection of labels. A dalograph is a finite directed graph whose arcs are labelled with elements of $L$, and such that every node has a non-zero outdegree. 
\end{defn}

The picture below is an example of dalograph, with 5 labels from $a_1$ to $a_5$. Squares represent nodes, but they may not be displayed in every further example.

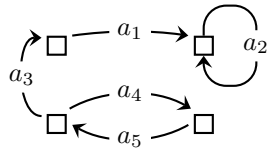
\[\psmatrix\\
  &[name=n1] \psframebox{} &\phantom{aaaaaaaa}& [name=n2] \psframebox{}\\\\
  &[name=n3] \psframebox{} && [name=n4] \psframebox{}\\
  \ncarc{->}{n1}{n2}
  \mput*{a_1}
  \ncloop[nodesep=0, arm=.4, angleA=90, angleB=-90, loopsize=.7, linearc=.3, arrows=->]{n2}{n2}
  \mput*{a_2}
  \ncarc[arcangleA=90, arcangleB=90]{->}{n3}{n1}
  \mput*{a_3}
  \ncarc[arcangleA=30, arcangleB=30]{->}{n3}{n4}
  \mput*{a_4}
  \ncarc[arcangleA=30, arcangleB=30]{->}{n4}{n3}  
  \mput*{a_5}
\endpsmatrix\]

One may argue that labelling the arcs leave out the node-labelled digraphs, as the example below.

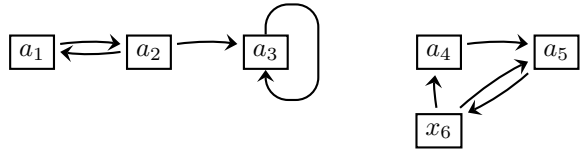
\[\psmatrix\\
      &[name=n1] \psframebox{a_1} &\phantom{aaaa}& [name=n2] \psframebox{a_2}&\phantom{aaaa}& [name=n3] \psframebox{a_3}&\phantom{aaaaaaaa}&
      [name=n4] \psframebox{a_4}&\phantom{aaaa}&[name=n5] \psframebox{a_5} \\\\
      &&&&&&&[name=n6] \psframebox{x_6}\\
      \ncarc{->}{n1}{n2}
      \ncarc{->}{n2}{n1}
      \ncarc{->}{n2}{n3}
      \ncloop[nodesep=0, arm=.4, angleA=90, angleB=-90, loopsize=.7, linearc=.3, arrows=->]{n3}{n3}
      \ncarc{->}{n4}{n5}
      \ncarc{->}{n5}{n6}
      \ncarc{->}{n6}{n5}
      \ncarc{->}{n6}{n4}
    \endpsmatrix\]

However, there is a natural way of embedding node-labelled digraphs into arc-labelled digraphs: for each node, remove the label from the node and put it on all the outgoing arcs of the node. The picture below shows the above example being translated into dalographs.

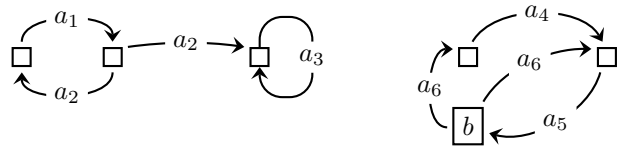
\[\psmatrix\\
      &[name=n1] \psframebox{} &\phantom{aaaa}& [name=n2] \psframebox{}&\phantom{aaaaaaaa}& [name=n3] \psframebox{}&\phantom{aaaaaaaaaaaa}&
      [name=n4] \psframebox{}&\phantom{aaaaaaa}&[name=n5] \psframebox{} \\\\
      &&&&&&&[name=n6] \psframebox{b}\\
      \ncarc[arcangleA=90, arcangleB=90]{->}{n1}{n2}
      \mput*{a_1}
      \ncarc[arcangleA=90, arcangleB=90]{->}{n2}{n1}  
      \mput*{a_2}
      \ncarc{->}{n2}{n3}
      \mput*{a_2}
      \ncloop[nodesep=0, arm=.4, angleA=90, angleB=-90, loopsize=.7, linearc=.3, arrows=->]{n3}{n3}
      \mput*{a_3}
      \ncarc[arcangleA=60, arcangleB=60]{->}{n4}{n5}
      \mput*{a_4}
      \ncarc[arcangleA=45, arcangleB=45]{->}{n5}{n6}
      \mput*{a_5}
      \ncarc[arcangleA=30, arcangleB=30]{->}{n6}{n5}
      \mput*{a_6}
      \ncarc[arcangleA=90, arcangleB=90]{->}{n6}{n4}
      \mput*{a_6}
    \endpsmatrix\]

Due to the embedding, arc labelling appears to be more general than node labelling.

Demanding that all nodes have non-zero outdegree may not yield a serious loss of generality either. Indeed there is also an embedding from arc-labelled digraphs into dalographs, using a dummy node and a dummy label $dl$. For example, this embedding maps the left-hand graph below to the right-hand dalograph.

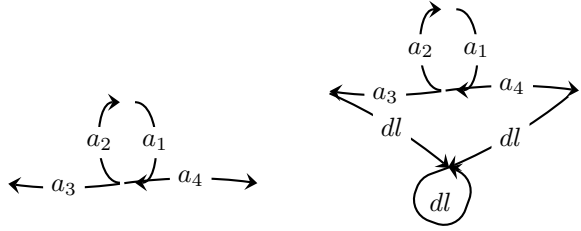
\[
\begin{array}{c@{\hspace{.8cm}}c}
\psmatrix
      &&&[name=n1]\\\\\\\\\\
      &[name=n2] &\phantom{aaaaaaaa}&[name=n3] &\phantom{aaaaaaaa}&[name=n4]\\\\
      \ncarc[nodesep=.1,arcangleA=90, arcangleB=90]{->}{n1}{n3}
      \mput*{a_1}
      \ncarc[nodesep=.1,arcangleA=90, arcangleB=90]{->}{n3}{n1}  
      \mput*{a_2}
      \ncarc[nodesep=.1]{->}{n3}{n2}
      \mput*{a_3}
      \ncarc[nodesep=.1]{->}{n3}{n4}
      \mput*{a_4}
    \endpsmatrix
&
\psmatrix
      &&&[name=n1]\\\\\\\\\\
      &[name=n2] &\phantom{aaaaaaaa}&[name=n3] &\phantom{aaaaaaaa}&[name=n4]\\\\\\\\\\
      &&&[name=n5]\\\\\\
      \ncarc[nodesep=.1,arcangleA=90, arcangleB=90]{->}{n1}{n3}
      \mput*{a_1}
      \ncarc[nodesep=.1,arcangleA=90, arcangleB=90]{->}{n3}{n1}  
      \mput*{a_2}
      \ncarc[nodesep=.1]{->}{n3}{n2}
      \mput*{a_3}
      \ncarc[nodesep=.1]{->}{n3}{n4}
      \mput*{a_4}
      \ncarc[nodesep=.1]{->}{n2}{n5}
      \mput*{dl}
      \ncarc[nodesep=.1]{->}{n4}{n5}
      \mput*{dl}
      \ncloop[nodesep=0, arm=.4, angleA=-160, angleB=-20, loopsize=.5, linearc=.3, arrows=->]{n5}{n5}
      \Aput{dl}
    \endpsmatrix
\end{array}\]

\subsection{Walks and Paths}\label{subsect:w-p}

This subsection defines a walk in a dalograph as a finite sequence of nodes that goes continuously along the arcs of the dalograph. In addition, a walk must stop when intersecting itself. A looping walk is defined as a walk intersecting itself. This enables the definition of an alternative induction principle along walks. Then, paths are defined as infinite sequences that are consistent with looping walks, and this definition is briefly justified. 

\begin{defn}[Walks as sequence of nodes]
Walks in a digraph are defined by induction as follows.
\begin{itemize}
\item $\epsilon$ is the empty walk.
\item $o$ is a walk for any node $o$ of the dalograph.
\item If $o_0\dots o_n$ is a walk, if $o$ does not occur in $o_
0,\dots,o_{n-1}$, and if $oo_0$ is an arc of the dalograph, $oo_0\dots o_n$ is also a walk.
\end{itemize}
If a node occurs twice in a walk, then it occurs at the end of the walk. In this case, the walk is said to be looping.
\end{defn}

The first picture below shows the walk $o_1o_2o_3$ using double lines. The second picture shows the looping walk $o_1o_2o_3o_2$.

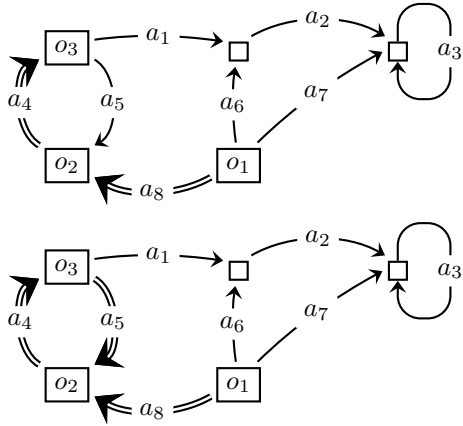
\[\psmatrix\\
   &[name=n1] \psframebox{o_3} &\phantom{aaaaaaaa}& [name=n2] \psframebox{}&\phantom{aaaaaaaa}& [name=n3] \psframebox{}\\\\\\
   &[name=n4] \psframebox{o_2}&\phantom{aaaaaaaa}&[name=n5] \psframebox{o_1}\\
   \ncarc[]{->}{n1}{n2}
   \mput*{a_1}
   \ncarc[arcangleA=30,arcangleB=30]{->}{n2}{n3}
   \mput*{a_2}
   \ncarc[arcangleA=60,arcangleB=60]{->}{n1}{n4}
   \mput*{a_5}
   \ncarc[arcangleA=60,arcangleB=60,doubleline=true]{->}{n4}{n1}
   \mput*{a_4}
   \ncarc[arcangleA=30,arcangleB=30,doubleline=true]{->}{n5}{n4}
   \mput*{a_8}
   \ncarc{->}{n5}{n2}
   \mput*{a_6}
   \ncarc{->}{n5}{n3}
   \mput*{a_7}
   \ncloop[nodesep=0, arm=.5, angleA=90, angleB=-90, loopsize=.7, linearc=.3, arrows=->]{n3}{n3}
   \mput*{a_3}
\endpsmatrix\]

\[\psmatrix
   &[name=n1] \psframebox{o_3} &\phantom{aaaaaaaa}& [name=n2] \psframebox{}&\phantom{aaaaaaaa}& [name=n3] \psframebox{}\\\\\\
   &[name=n4] \psframebox{o_2}&\phantom{aaaaaaaa}&[name=n5] \psframebox{o_1}\\
   \ncarc[]{->}{n1}{n2}
   \mput*{a_1}
   \ncarc[arcangleA=30,arcangleB=30]{->}{n2}{n3}
   \mput*{a_2}
   \ncarc[arcangleA=60,arcangleB=60,doubleline=true]{->}{n1}{n4}
   \mput*{a_5}
   \ncarc[arcangleA=60,arcangleB=60,doubleline=true]{->}{n4}{n1}
   \mput*{a_4}
   \ncarc[arcangleA=30,arcangleB=30,doubleline=true]{->}{n5}{n4}
   \mput*{a_8}
   \ncarc{->}{n5}{n2}
   \mput*{a_6}
   \ncarc{->}{n5}{n3}
   \mput*{a_7}
   \ncloop[nodesep=0, arm=.5, angleA=90, angleB=-90, loopsize=.7, linearc=.3, arrows=->]{n3}{n3}
   \mput*{a_3}
\endpsmatrix\]

It will often be convenient to represent walks partially. A (possibly empty) non-looping walk is represented by a double-headed arrow that is labelled with a sequence of nodes. For instance, the left-hand walk $u=u_1\dots u_n$ below is represented by the right-hand picture.

\[\psmatrix\\
    &[name=n3]\psframebox{u_1}& \phantom{aaaaaa}&[name=n4] \psframebox{\phantom{aa}} &\phantom{aa}\dots\phantom{aa}&[name=n5] \psframebox{\phantom{aa}}&\phantom{aaaaaa}&[name=n6]\psframebox{u_n}&\phantom{aaaaaaaa}&[name=n1]&\phantom{aaaaaaaa}&[name=n2]\\
    \ncarc{->>}{n1}{n2}
    \mput*{u}
    \ncarc{->}{n3}{n4}
    \ncarc{->}{n5}{n6}
\endpsmatrix\]

A looping walk is represented by a squared-bracket-headed arrow that is labelled with a sequence of nodes. For instance below, the left-hand looping walk $uovo$ is represented by the right-hand picture.\\ 

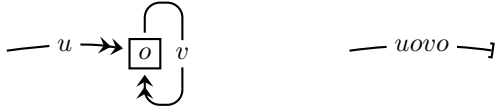
\[\psmatrix
    &[name=n1] &\phantom{aaaaaaaa}&[name=n2] \psframebox{o}&\phantom{aaaaaaaaaaaa}&[name=n] &\phantom{aaaaaaaaaa}&[name=n'] \\\\
    \ncarc{-]}{n}{n'}
    \mput*{uovo}
    \ncarc{->>}{n1}{n2}
    \mput*{u}
    \ncloop[arm=.4, angleA=90, angleB=-90, loopsize=.5, linearc=.2, arrows=->>]{n2}{n2}
    \mput*{v}
\endpsmatrix\]

A usual induction principal for walks would go from the empty walk to bigger walks. Here, walks take place in finite dalographs so they are bounded. This allows an alternative induction principle for walks. 

\begin{lem}[Nibbling induction principle for walks]
Let $g$ be a dalograph. Let $P$ be a predicate on walks in $g$. Assume that given any walk $x$, $P(xo)$ for all walks $xo$ implies $P(x)$. Then the predicate holds for all walks.
\end{lem}

\begin{proof}
First, $P$ holds for all looping walks since, by definition, $x$ being a looping walk implies $xo$ is not a walk. Second, assume that there exist walks that do not satisfy $P$. Let $x$ be one such walk. By finiteness of $g$, it makes sense to take $x$ as long as possible. So all walks $xo$ satisfy $P$ by definition of $x$, therefore $x$ satisfies $P$ by assumption. This is a contradiction.
\end{proof}

This lemma can be used as a proof principle or, quite similarly, as a programming principle. Indeed, let $f$ be a function on walks. If the definition of $f$ on any input $x$ invokes only results of the computation of $f$ with the walks $xo$, then $f$ is well-defined a priori. 

Paths are defined as consistent infinite continuations of looping walks.

\begin{defn}[Paths as looping walks]
Given a looping walk $uovo$, the corresponding path is the infinite sequence $u(ov)^\omega$. Given a walk $x$ and a path $\Gamma$ such that $x\Gamma$ is also a path, $\Gamma$ is called a continuation of the walk $x$.
\end{defn}

Informally, a path has a memory, and when visiting a node for the second time and more, it chooses the same next node as for the first time. The formalism ensures the following properties.

\begin{lem}
Every walk has a continuation and a looping walk $uovo$ has a unique continuation $(vo)^\omega$.
\end{lem}

If one wishes to deal with "finite paths", \textit{i.e.} non-looping walks, within the dalograph formalism, it is possible to add a dummy node and dummy arcs from all original nodes to the dummy one, as shown below.\\

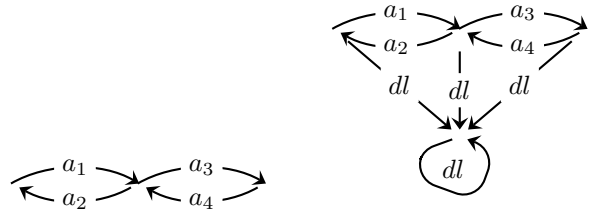
\[
\begin{array}{c@{\hspace{.8cm}}c}
\psmatrix
      &[name=n1] &\phantom{aaaaaaaa}&[name=n2] &\phantom{aaaaaaaa}&[name=n3]\\\\
      \ncarc[nodesep=.1,arcangleA=30, arcangleB=30]{->}{n1}{n2}
      \mput*{a_1}
      \ncarc[nodesep=.1,arcangleA=30, arcangleB=30]{->}{n2}{n1}  
      \mput*{a_2}
      \ncarc[nodesep=.1,arcangleA=30, arcangleB=30]{->}{n2}{n3}  
      \mput*{a_3}
      \ncarc[nodesep=.1,arcangleA=30, arcangleB=30]{->}{n3}{n2} 
      \mput*{a_4}
    \endpsmatrix
&
\psmatrix
      &[name=n1] &\phantom{aaaaaaaa}&[name=n2] &\phantom{aaaaaaaa}&[name=n3]\\\\\\\\\\\\\\
      &&&[name=n4]\\\\\\\\\\
      \ncarc[nodesep=.1,arcangleA=30, arcangleB=30]{->}{n1}{n2}
      \mput*{a_1}
      \ncarc[nodesep=.1,arcangleA=30, arcangleB=30]{->}{n2}{n1}  
      \mput*{a_2}
      \ncarc[nodesep=.1,arcangleA=30, arcangleB=30]{->}{n2}{n3}  
      \mput*{a_3}
      \ncarc[nodesep=.1,arcangleA=30, arcangleB=30]{->}{n3}{n2} 
      \mput*{a_4}
      \ncline[nodesep=.2]{->}{n1}{n4}
      \mput*{dl}
      \ncline[nodesep=.2]{->}{n2}{n4}
      \mput*{dl}
      \ncline[nodesep=.2]{->}{n3}{n4}
      \mput*{dl}
      \ncloop[nodesep=.1, arm=.4, angleA=-160, angleB=-20, loopsize=.4, linearc=.3, arrows=->]{n4}{n4}
      \Aput{dl}
    \endpsmatrix
\end{array}\]

Through the embedding above, the left-hand infinite path below may be interpreted as the right-hand finite path.\\

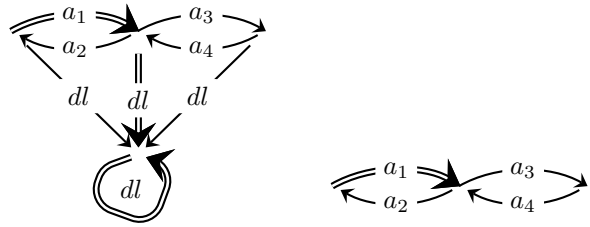
\[
\begin{array}{c@{\hspace{.8cm}}c}
\psmatrix
      &[name=n1] &\phantom{aaaaaaaa}&[name=n2] &\phantom{aaaaaaaa}&[name=n3]\\\\\\\\\\\\\\\\
      &&&[name=n4]\\\\\\\\\\
      \ncarc[doubleline=true,nodesep=.1,arcangleA=30, arcangleB=30]{->}{n1}{n2}
      \mput*{a_1}
      \ncarc[nodesep=.1,arcangleA=30, arcangleB=30]{->}{n2}{n1}  
      \mput*{a_2}
      \ncarc[nodesep=.1,arcangleA=30, arcangleB=30]{->}{n2}{n3}  
      \mput*{a_3}
      \ncarc[nodesep=.1,arcangleA=30, arcangleB=30]{->}{n3}{n2} 
      \mput*{a_4}
      \ncline[nodesep=.2]{->}{n1}{n4}
      \mput*{dl}
      \ncline[nodesep=.2,doubleline=true]{->}{n2}{n4}
      \mput*{dl}
      \ncline[nodesep=.2]{->}{n3}{n4}
      \mput*{dl}
      \ncloop[doubleline=true,nodesep=.1, arm=.4, angleA=-160, angleB=-20, loopsize=.5, linearc=.3, arrows=->]{n4}{n4}
      \Aput{dl}
    \endpsmatrix
&
\psmatrix
      &[name=n1] &\phantom{aaaaaaaa}&[name=n2] &\phantom{aaaaaaaa}&[name=n3]\\\\
      \ncarc[doubleline=true,nodesep=.1,arcangleA=30, arcangleB=30]{->}{n1}{n2}\\
      \mput*{a_1}
      \ncarc[nodesep=.1,arcangleA=30, arcangleB=30]{->}{n2}{n1}  
      \mput*{a_2}
      \ncarc[nodesep=.1,arcangleA=30, arcangleB=30]{->}{n2}{n3}  
      \mput*{a_3}
      \ncarc[nodesep=.1,arcangleA=30, arcangleB=30]{->}{n3}{n2} 
      \mput*{a_4}
    \endpsmatrix
\end{array}\]

\subsection{Equilibria}\label{subsect:eq}

This subsection compares paths by comparing infinite sequences of labels. It defines strategies as dalographs having chosen an outgoing arc for each of their nodes, and it defines equilibria as strategies inducing maximal paths everywhere. 

Below, finite sequences of labels induced by walks are defined by induction (the usual induction principle, not the nibbling one). Subsequently, paths induce infinite sequences of labels.

\begin{defn}[Induced sequence]
Induced sequences are inductively defined as follows.
\begin{itemize}
\item A walk $o$ induces the empty sequence.
\item A walk $o_1o_2x$ induces the sequence $a.seq(o_2x)$ where $a$ is the label on the arc $o_1o_2$.
\end{itemize}
A path corresponding to a looping walk $uovo$ induces a sequence $seq(uo)seq(ovo)^\omega$. A sequence that is induced by a path starting from a given node is said eligible at this node.
\end{defn}

The ultimately periodic sequences over labels are comparable through an arbitrary binary relation.

\begin{defn}[Preference]
A relation over the ultimately periodic sequences over labels is called a preference. Preferences are written $\prec$ in this chapter.
\end{defn}

Comparing the induced sequences of two paths starting from the same node is a natural way of comparing paths.

\begin{defn}[Paths comparison]
If paths $\Gamma_1$ and $\Gamma_2$ start from the same node and induce sequences $\gamma_1$ and $\gamma_2$ with $\gamma_1\prec\gamma_2$ (resp. $\gamma_1\not\prec\gamma_2$), one writes $\Gamma_1\prec\Gamma_2$ (resp. $\Gamma_1\not\prec\Gamma_2$) by abuse of notation.
\end{defn}

The following definition captures the notion of maximality (with respect to a preference) of a path among the continuations of a given walk.

\begin{defn}[Maximal continuation]
The notation $m_{g,\prec}(xo,\Gamma)$ accounts for the property: $xo\Gamma$ is a path in $g$ and $o\Gamma\not\prec o\Gamma'$ for all paths $xo\Gamma'$ of the dalograph $g$.
\end{defn}

One may write $m(xo,\Gamma)$ instead of $m_{g,\prec}(xo,\Gamma)$ when there is no ambiguity. In addition to a path being maximal from the point of view of its starting node, this chapter needs to discuss paths all of whose subpaths are maximal from their starting points. The notion of hereditary maximality captures this idea below.

\begin{defn}[Hereditary maximal path]
Let $\Gamma$ be a path. If $m_{g,\prec}(o,\Gamma')$ for any decomposition $\Gamma=xo\Gamma'$ where $xo$ is a non-looping walk, one writes $hm_{g,\prec}(\Gamma)$.
\end{defn}

A strategy is an object built on a dalograph by choosing an outgoing arc at each node.

\begin{defn}[Strategy]
Given a dalograph $g$, a strategy $s$ on $g$ is a pair $(g,c)$, where $c$ is a function from the nodes of $g$ to themselves, and such that for all nodes $o$, the pair $(o,c(o))$ is an arc of $g$.
\end{defn}

The two examples below show two strategies with the same underlying dalograph. The choices are represented by double lines.\\

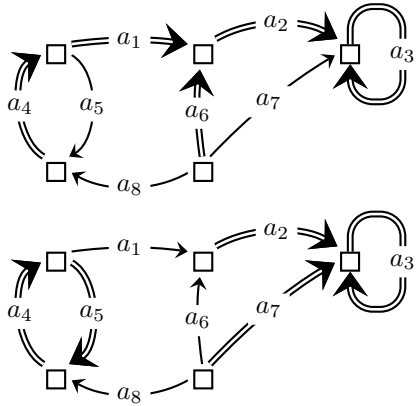
\[\psmatrix
   &[name=n1] \psframebox{} &\phantom{aaaaaaaa}& [name=n2] \psframebox{}&\phantom{aaaaaaaa}& [name=n3] \psframebox{}\\\\\\
   &[name=n4] \psframebox{}&\phantom{aaaaaaaa}&[name=n5] \psframebox{}\\
   \ncarc[doubleline=true]{->}{n1}{n2}
   \mput*{a_1}
   \ncarc[arcangleA=30,arcangleB=30,doubleline=true]{->}{n2}{n3}
   \mput*{a_2}
   \ncarc[arcangleA=60,arcangleB=60]{->}{n1}{n4}
   \mput*{a_5}
   \ncarc[arcangleA=60,arcangleB=60,doubleline=true]{->}{n4}{n1}
   \mput*{a_4}
   \ncarc[arcangleA=30,arcangleB=30]{->}{n5}{n4}
   \mput*{a_8}
   \ncarc[doubleline=true]{->}{n5}{n2}
   \mput*{a_6}
   \ncarc{->}{n5}{n3}
   \mput*{a_7}
   \ncloop[doubleline=true, nodesep=0, arm=.5, angleA=90, angleB=-90, loopsize=.7, linearc=.3, arrows=->]{n3}{n3}
   \mput*{a_3}
\endpsmatrix\]

\[\psmatrix
   &[name=n1] \psframebox{} &\phantom{aaaaaaaa}& [name=n2] \psframebox{}&\phantom{aaaaaaaa}& [name=n3] \psframebox{}\\\\\\
   &[name=n4] \psframebox{}&\phantom{aaaaaaaa}&[name=n5] \psframebox{}\\
   \ncarc{->}{n1}{n2}
   \mput*{a_1}
   \ncarc[arcangleA=30,arcangleB=30,doubleline=true]{->}{n2}{n3}
   \mput*{a_2}
   \ncarc[doubleline=true,arcangleA=60,arcangleB=60]{->}{n1}{n4}
   \mput*{a_5}
   \ncarc[arcangleA=60,arcangleB=60,doubleline=true]{->}{n4}{n1}
   \mput*{a_4}
   \ncarc[arcangleA=30,arcangleB=30]{->}{n5}{n4}
   \mput*{a_8}
   \ncarc{->}{n5}{n2}
   \mput*{a_6}
   \ncarc[doubleline=true]{->}{n5}{n3}
   \mput*{a_7}
   \ncloop[doubleline=true, nodesep=0, arm=.5, angleA=90, angleB=-90, loopsize=.7, linearc=.3, arrows=->]{n3}{n3}
   \mput*{a_3}
\endpsmatrix\]

\medskip

As seen in the pictures above, given a strategy and a node, the strategy induces exactly one path starting from this node. 

\begin{defn}[Induced continuation]
Let $s$ be a strategy. Define $p(s,o)$ such that $o\cdot p(s,o)$ is the path induced by $s$ starting from $o$.
\end{defn}

Given a dalograph and a preference, a strategy on the dalograph is a local equilibrium at a given node if it induces a hereditary maximal path at this node.

\begin{defn}[Local equilibrium]
$LEq_{\prec}(s,o)\,\eqdef\,m_{g,\prec}(o,p(s,o))$ 
\end{defn}

A global equilibrium for a dalograph is intended to be a strategy inducing a maximal path at every node of the dalograph. It follows that a global equilibrium can be defined as a strategy that is a local equilibrium for every node of the dalograph.

\begin{defn}[Global equilibrium]
$GEq_{\prec}(s)\,\eqdef\,\forall o\in g,\,LEq_{\prec}(s,o)$ 
\end{defn}

In the rest of this chapter, the terminology of equilibrium refers to global equilibria, unless otherwise stated, or $\prec$-equilibrium to avoid ambiguity. In the example below, arcs are labelled with natural numbers and $\prec$ is the lexicographic extension of the usual order to infinite sequences of natural numbers. The following strategy is a local equilibrium for node $o'$ but not for node $o$.

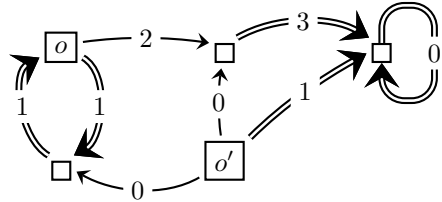
\[\psmatrix\\
   &[name=n1] \psframebox{o} &\phantom{aaaaaaaa}& [name=n2] \psframebox{}&\phantom{aaaaaaaa}& [name=n3] \psframebox{}\\\\\\
   &[name=n4] \psframebox{}&\phantom{aaaaaaaa}&[name=n5] \psframebox{o'}\\
   \ncarc{->}{n1}{n2}
   \mput*{2}
   \ncarc[arcangleA=30,arcangleB=30,doubleline=true]{->}{n2}{n3}
   \mput*{3}
   \ncarc[doubleline=true,arcangleA=60,arcangleB=60]{->}{n1}{n4}
   \mput*{1}
   \ncarc[arcangleA=60,arcangleB=60,doubleline=true]{->}{n4}{n1}
   \mput*{1}
   \ncarc[arcangleA=30,arcangleB=30]{->}{n5}{n4}
   \mput*{0}
   \ncarc{->}{n5}{n2}
   \mput*{0}
   \ncarc[doubleline=true]{->}{n5}{n3}
   \mput*{1}
   \ncloop[doubleline=true, nodesep=0, arm=.5, angleA=90, angleB=-90, loopsize=.7, linearc=.3, arrows=->]{n3}{n3}
   \mput*{0}
\endpsmatrix\]

If a preference is a subrelation of another preference, and if a given strategy is a local/global equilibrium with respect to the bigger preference, then the strategy is also a local/global equilibrium with respect to the smaller preference. This is formally stated below.

\begin{lem}[Equilibrium for subpreference]\label{lem:dalo-esp} Preservation by subrelation is stated as follows.
\[\prec\subseteq\prec'\,\Rightarrow\,LEq_{\prec'}(s,o)\,\Rightarrow\,LEq_{\prec}(s,o)\]
\[\prec\subseteq\prec'\,\Rightarrow\,GEq_{\prec'}(s,o)\,\Rightarrow\,GEq_{\prec}(s,o)\]
\end{lem}

\begin{proof}
Note that the following formula holds.\\ $\prec\subseteq\prec'\,\Rightarrow\,m_{g,\prec'}(xo,\Gamma)\,\Rightarrow\,m_{g,\prec}(xo,\Gamma)$.
\end{proof}

Assume that two preferences coincide on a subdomain, \textit{i.e.} a subset of the ultimately periodic sequences over labels. Assume that a given dalograph involves only sequences from this subdomain, \textit{i.e.} all paths in the dalograph induce only sequences in the subdomain. In this case, a local/global equilibrium for this dalograph with respect to one preference is also a local/global equilibrium for this dalograph with respect to other preference. As for the lemma above, this can be proved by simple unfolding of the definitions. This result is stated below.

\begin{lem}\label{lem:coincide-eq}
Let $g$ be a dalograph involving sequences in $S$ only. Assume that $\prec|_S=\prec'|_S$. In this case, 
\[LEq_{\prec}(s,o)\quad\Leftrightarrow\quad LEq_{\prec'}(s,o)\]
\[GEq_{\prec}(s)\quad\Leftrightarrow\quad GEq_{\prec'}(s)\]
\end{lem}

The following lemma relates to the fact that when there is only one choice, this choice is the best possible one.

\begin{lem}\label{lem:no-choice-eq}
If $\prec$ is an irreflexive preference, then a dalograph each of whose node has outdegree $1$ has a $\prec$-equilibrium.
\end{lem}

\begin{proof}
In such a case, only one strategy corresponds to that dalograph and there is only one possible path starting from any node. Therefore, each path is maximal, by irreflexivity, and this strategy is an equilibrium.
\end{proof}

\section{Binary Relations}\label{sect:br}

This section defines a few predicates on binary relations. Some properties connecting these predicates are also presented. Subsection~\ref{subsect:gbr} deals with binary relations in general, while subsection~\ref{subsect:brs} focuses on binary relations over sequences.

\subsection{General Binary Relations}\label{subsect:gbr}

This subsection slightly rephrases the notion of strict weak order, which already exists in the literature. This subsection also defines the notion of imitation. It turns out that strict weak orders can be equivalently characterised by a few other simple formulae. The structure of such relations is studied in detail.

The notions of transitive, asymmetric, and irreflexive binary relation that are used in this chapter are the usual ones.

\begin{defn}[Transitivity, asymmetry and irreflexivity]
A binary relation $\prec$ is transitive if it complies with the first formula below. Asymmetry amounts to the second formula, and irreflexivity to the third one.
\[\begin{array}{c@{\qquad}l}
\alpha\prec\beta\,\Rightarrow\,\beta\prec\gamma\,\Rightarrow\,\alpha\prec\gamma & transitivity\\
\alpha\prec\beta\,\Rightarrow\,\beta\not\prec\alpha & asymmetry\\
\alpha\not\prec\alpha & irreflexivity
\end{array}\]
\end{defn}

Asymmetry implies irreflexivity, as being formally stated below.

\begin{lem}\label{lem:a-i}
$(\forall\alpha,\beta,\,\alpha\prec\beta\,\Rightarrow\,\beta\not\prec\alpha)\,\Rightarrow\,\forall\alpha,\alpha\not\prec\alpha$
\end{lem}

\begin{proof}
Instantiate $\beta$ with $\alpha$.
\end{proof}

Transitivity of the negation of a relation does not imply transitivity of the relation. For instance, let $\alpha\neq\beta$ and define $\prec$ on $\{\alpha,\beta\}$ by $\alpha\prec\beta$ and $\beta\prec\alpha$, and nothing more. Due to the symmetry, $\prec$ is not transitive, while $\not\prec$ is transitive. The next definition and two lemmas shows that transitivity of the negation almost implies transitivity.

\begin{defn}[Strict weak order]
A strict weak order is an asymmetric relation whose negation is transitive.
\end{defn}

Equivalent definitions of strict weak order can be found in the literature. The rest of this subsection explains some properties of strict weak orders and it gives an intuition of the underlying structure. The following lemma shows that transitivity of a relation can be derived from asymmetry of the relation and transitivity of its negation.

\begin{lem}\label{lem:act-t}
A strict weak order is transitive. 
\end{lem}

\begin{proof}
Let $\alpha$, $\beta$ and $\gamma$ be such that $\alpha\prec\beta$ and $\beta\prec\gamma$. Therefore $\beta\not\prec\alpha$ and $\gamma\not\prec\beta$ by asymmetry, and  $\gamma\not\prec\alpha$ by transitivity of the negation . If $\alpha\not\prec\gamma$ then $\beta\not\prec\alpha$ and transitivity of the negation  yields $\beta\not\prec\gamma$, which is absurd. Therefore $\alpha\prec\gamma$.
\end{proof}

Strict weak orders have a second property that makes non-comparability an equivalence relation.

\begin{lem}\label{lem:kr-eq}
If $\prec$ is a strict weak order, then $\sharp$ is an equivalence relation.
\end{lem}

\begin{proof}
If $\neg(\alpha\sharp\alpha)$ then $\alpha\prec\alpha$, which contradicts asymmetry by lemma~\ref{lem:a-i}. So $\sharp$ is reflexive. If $\alpha\sharp\beta$ then $\beta\sharp\alpha$ by definition, so $\sharp$ is symmetric. If $\alpha\sharp\beta$ and $\beta\sharp\gamma$, then $\beta\not\prec\alpha$ and $\gamma\not\prec\beta$ by definition. So $\gamma\not\prec\alpha$ by transitivity of the negation. In the same way, $\alpha\not\prec\gamma$ also holds, so $\alpha\sharp\gamma$. Therefore $\sharp$ is transitive. The incomparability relation $\sharp$ is symmetric by definition, so $\sharp$ is an equivalence relation.
\end{proof}

A binary relation is lower (resp. upper) imitating if any two non-comparable elements have the same predecessors (resp. successors).

\begin{defn}[Lower/upper imitation]
A binary relation $\prec$ complying with the following formula is called a lower-imitating relation.
 \[(\alpha\sharp\beta\,\wedge\,\gamma\prec\alpha)\Rightarrow\gamma\prec\beta\]
A binary relation $\prec$ complying with the following formula is called a upper-imitating relation.
\[(\alpha\sharp\beta\,\wedge\,\alpha\prec\gamma)\Rightarrow\beta\prec\gamma\] 
A relation that is both lower and upper imitating is called an imitating relation.
\end{defn}

Lower and upper imitations do not only look "symmetric" definitions, they also are.

\begin{lem}\label{lem:luis}
If a relation is lower (resp. upper) imitating, then its inverse is upper (resp. lower) imitating.
\end{lem}

\begin{proof}
Let $\prec$ be a binary relation. Assume that $\prec$ is lower imitating and assume that $\alpha\sharp^{-1}\beta$ and $\beta\succ\gamma$. So $\alpha\sharp\beta$ and $\gamma\prec\beta$, which implies $\gamma\prec\alpha$, and therefore $\alpha\succ\gamma$. 
\end{proof}

Guaranteeing asymmetry makes the predicates of lower and upper imitations coincide.

\begin{lem}\label{lem:a-lui}
Let $\prec$ be an asymmetric relation. In this case, $\prec$ is upper-imitating \textit{iff} $\prec$ is lower-imitating.
\end{lem}

\begin{proof}
left-to-right. Assume $\alpha\sharp\beta$ and $\gamma\prec\beta$. If $\alpha\prec\gamma$ then $\beta\prec\gamma$ by upper-imitation, which contradicts asymmetry. If $\alpha\sharp\gamma$ then $\alpha\prec\beta$ by upper-imitation, which is absurd. So $\gamma\prec\alpha$. The converse follows lemma~\ref{lem:luis}, knowing that the inverse of an asymmetric relation is asymmetric.
\end{proof}

Because of the lemma above, only the concept of imitation will be referred to when dealing with asymmetric relations. The next lemma connects the notion of imitation to the notion of transitivity.

\begin{lem}\label{lem:teq-i}
If $\prec$ is transitive and $\sharp$ is an equivalence relation, then $\prec$ is irreflexive and imitating.
\end{lem}

\begin{proof}
If $\alpha\prec\alpha$ then $\sharp$ is not reflexive, which contradicts the assumption. So $\prec$ is irreflexive. Assume that $\alpha\sharp\beta$ and $\beta\prec\gamma$. If $\gamma\prec\alpha$ then $\alpha\prec\beta$ by transitivity, which is absurd. If $\gamma\sharp\alpha$, then $\beta\sharp\gamma$ by transitivity of $\sharp$, which is absurd. Therefore $\alpha\prec\gamma$.
\end{proof}

Imitation implies transitivity provided that "small" cycles are forbidden.

\begin{lem}\label{lem:nci-t}
Let $\prec$ be without any cycle involving 2 or 3 elements. If $\prec$ is imitating then $\prec$ is transitive. 
\end{lem}

\begin{proof}
Assume $\alpha\prec\beta$ and $\beta\prec\gamma$. If $\gamma\prec\alpha$ then there is a cycle involving 3 elements, which is absurd. Now assume that $\gamma\sharp\alpha$. If $\prec$ is imitating then $\beta\prec\alpha$, which contradicts asymmetry. So $\alpha\prec\gamma$. 
\end{proof}

Actually, an imitating relation is a strict weak order provided that "small" cycles are forbidden.

\begin{lem}\label{lem:ti-ct}
Let $\prec$ be without any cycle of length 2 or 3. If $\prec$ is imitating then $\prec$ is a strict weak order. 
\end{lem}

\begin{proof}
Let $\alpha$, $\beta$ and $\gamma$ be such that $\beta\not\prec\alpha$ and $\gamma\not\prec\beta$. First case, assume that $\alpha\sharp\beta$ and $\beta\sharp\gamma$. If $\gamma\prec\alpha$ then $\beta\prec\alpha$ and $\gamma\prec\beta$ by double imitation. Since transitivity is guaranteed by lemma~\ref{lem:nci-t}, this yields $\gamma\prec\alpha$ which is absurd by assumption. Second case, either $\alpha\sharp\beta$ and $\beta\prec\gamma$ or $\alpha\prec\beta$ and $\beta\sharp\gamma$. By imitation, $\alpha\prec\gamma$, so $\gamma\not\prec\alpha$ by asymmetry. Third case, $\alpha\prec\beta$ and $\beta\prec\gamma$, so $\alpha\prec\gamma$ by transitivity, so $\gamma\not\prec\alpha$.
\end{proof}

The picture below is meant to give an intuitive understanding of what is a strict weak order. The circles represent equivalence classes of $\sharp$. Here we have $\gamma\prec\alpha,\beta,\delta$ and $\gamma,\alpha,\beta\prec\delta$ and $\alpha\sharp\beta$. Informally, a strict weak order looks like a knotted rope.

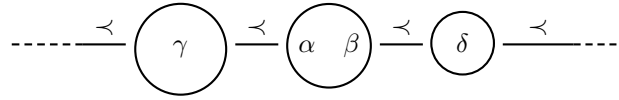
\[\psmatrix
  \\
  &[name=n1] &\phantom{aaaa}& [name=n2] &\phantom{aa}& [name=n3] & [name=n4]\pscirclebox{\phantom{aa}\gamma\phantom{aa}}
  &[name=n5] &\phantom{aa}& [name=n6] & [name=n7]\pscirclebox{\alpha\phantom{aa}\beta}
  &[name=n8] &\phantom{aa}& [name=n9] & [name=n10]\pscirclebox{\phantom{a}\delta\phantom{a}}
  &[name=n11] &\phantom{aaaa}& [name=n12] &\phantom{aa}& [name=n13]
  \\
  \ncline[nodesep=0,linestyle=dashed,dash=3pt 2pt]{-}{n1}{n2}
  \ncline[nodesep=0]{-}{n2}{n3}
  \Aput{\prec}
  \ncline[nodesep=0]{-}{n5}{n6}
  \Aput{\prec}
  \ncline[nodesep=0]{-}{n8}{n9}
  \Aput{\prec}
  \ncline[nodesep=0]{-}{n11}{n12}
  \Aput{\prec}
  \ncline[nodesep=0,linestyle=dashed,dash=3pt 2pt]{-}{n12}{n13}
  \endpsmatrix\]

The following lemma sums up the results of this subsection.

\begin{lem}\label{lem:kn-teq-ic}
Let $\prec$ be a binary relation, and let $\sharp$ be the corresponding non-comparability relation. The following three propositions are equivalent.
\begin{enumerate}
\item $\prec$ is a strict weak order.
\item $\prec$ is transitive and $\sharp$ is an equivalence relation.
\item $\prec$ is imitating and has no cycle of length 2 or 3.
\end{enumerate}
\end{lem}

\begin{proof}
Implication $1\,\Rightarrow\,2$ by lemmas~\ref{lem:act-t} and~\ref{lem:kr-eq}, implication $2\,\Rightarrow\,3$ by lemma~\ref{lem:teq-i}, and implication $3\,\Rightarrow\,1$ by lemma~\ref
{lem:ti-ct}.
\end{proof}

\subsection{Binary Relations over Sequences}\label{subsect:brs}

This subsection deals with binary relations over (finite or infinite) sequences built over (finite or infinite) collections. It introduces a few notions such as E-prefix, A-transitivity and subcontinuity.

For binary relations over sequences, the following captures the notions of preservation by prefix elimination and preservation by prefix addition.

\begin{defn}[E-prefix and A-prefix]
A binary relation $\prec$ over sequences is said E-prefix when complying with the following formula.
\[u\alpha\prec u\beta\,\Rightarrow\,\alpha\prec\beta\]
It is said A-prefix when complying with the following formula.
\[\alpha\prec \beta\,\Rightarrow\,u\alpha\prec u\beta\]
\end{defn}

It is possible to define a mix between transitivity and A-prefix.

\begin{defn}[A-transitivity]
A relation over sequences is said A-transitive when complying with the following formula.
\[\alpha\prec\beta\,\Rightarrow\,u\beta\prec\gamma\,\Rightarrow\,u\alpha\prec\gamma\]
\end{defn}

The following lemma shows the connections between transitivity, A-prefix, and A-transitivity. Note that the converse implications do not hold a priori.

\begin{lem}
Transitivity plus A-prefix imply A-transitivity, and A-transitivity implies transitivity.
\end{lem}

\begin{proof}
Consider the formula defining A-transitivity. For the first claim, $u\alpha\prec u\beta$ by A-prefix, and conclude by transitivity. For the second claim, instantiate $u$ with the empty sequence. 
\end{proof}

The following lemma shows that a strict weak order that is preserved by prefix elimination is A-transitive.

\begin{lem}\label{lem:Ekr-ut}
An E-prefix strict weak order is A-transitive.
\end{lem}

\begin{proof}
Assume that $\alpha\prec\beta$ and $u\beta\prec\gamma$. Therefore $\beta\not\prec\alpha$ by asymmetry, so $u\beta\not\prec u\alpha$ by contraposition of E-prefix. If $u\alpha\not\prec \gamma$ then $u\beta\not\prec\gamma$ by transitivity of the negation, which contradicts the assumption. Therefore $u\alpha\prec\gamma$. 
\end{proof}

The A-prefix predicate seems to be a bit too restrictive for what is intended in this chapter, but a somewhat related notion will be useful. Informally, consider a relation $\prec$ that is A-prefix and transitive. If $\alpha\prec u\alpha$ then $u\alpha\prec u^2\alpha$ and we have an infinite ascending chain $\alpha\prec u\alpha\prec\dots\prec u^n\alpha\prec\dots$. In this case, a natural thought might be to topologically close the chain with $u^\omega$ as an upper bound, \textit{i.e.} $\alpha\prec u\alpha\prec\dots\prec u^n\alpha\prec\dots\prec u^\omega$. The following definition captures this informal thought.

\begin{defn}[Subcontinuity]
A relation over sequences is said subcontinuous when complying with the following formula, where $u$ is any non-empty finite sequence.
\[\alpha\prec u\alpha\quad\Rightarrow\quad u^\omega\not\prec\alpha\]
\end{defn}

The next definition proceeds in the vein of the previous one and gives an alternative, slightly more complex definition of subcontinuity. 

\begin{defn}[Alt-subcontinuity]
A relation over sequences is alt-subcontinuous when complying with the following formula, where $v$ and $t$ are any non-empty finite sequences.
\[\alpha\prec t\beta\,\Rightarrow\, (v\alpha\prec \beta\,\vee\,\alpha\prec(tv)^\omega)\]
\end{defn}

The next lemma shows that alt-subcontinuity is "stronger" than subcontinuity.

\begin{lem}
An alt-subcontinuous asymmetric relation is subcontinuous.
\end{lem}

\begin{proof}
Let $\prec$ be an alt-subcontinuous asymmetric relation. Assume that $\alpha\prec u\alpha$ for some $u$ and $\alpha$. By alt-subcontinuity, $u\alpha\prec\alpha\,\vee\,\alpha\prec u^\omega$. By asymmetry, $u\alpha\not\prec\alpha$, so $\alpha\prec u^\omega$. Therefore $u^\omega\not\prec\alpha$ by asymmetry.
\end{proof}

The following lemma states that, under some demanding conditions, subcontinuity implies alt-subcontinuity.

\begin{lem}\label{lem:aESc}
A E-prefix subcontinuous strict weak order is alt-subcontinuous.
\end{lem}

\begin{proof}
Assume that $v\alpha\not\prec \beta$, $\alpha\not\prec (tv)^\omega$, and $\alpha\prec\ t\beta$. If $(tv)^\omega\not\prec t\beta$ then $\alpha\not\prec\ t\beta$ by transitivity of $\not\prec$, which is a contradiction, so $(tv)^\omega\prec t\beta$. So $(vt)^\omega\prec\beta$ by E-prefix. If $(vt)^\omega\not\prec v\alpha$ then $(vt)^\omega\not\prec\beta$ by transitivity of $\not\prec$, which is a contradiction, so $(vt)^\omega\prec v\alpha$, and $(tv)^\omega\prec\alpha$. By subcontinuity, this implies $\alpha\not\prec tv\alpha$. By assumption, $v\alpha\not\prec \beta$, so $tv\alpha\not\prec t\beta$ by contraposition of E-prefix. Therefore $\alpha\not\prec t\beta$ by transitivity of the negation, which contradicts the assumption, so $\alpha\not\prec t\beta$.
\end{proof}

The following two definitions generalise the notion of E-prefix.

\begin{defn}
Let $W$ be a non-empty finite set of finite sequences such that the empty sequence is not in $W$ and such that at most one sequence of length 1 is in $W$. If for all walks $u$ in $W$ there exists a walk $v$ in $W$ such that $u\alpha\prec v\beta$, one writes $W\alpha\prec W\beta$.
\end{defn}

\begin{defn}[Gen-E-prefix]
A relation over sequences is said gen-E-prefix when complying with the following formula.
\[W\alpha\prec W\beta\,\Rightarrow\,\alpha\prec\beta\]
\end{defn}

\section{Equilibrium Existence}\label{sect:eq-ex}

Subsection~\ref{subsect:p-eq-ex} proves a sufficient condition that guarantees existence of equilibrium in any dalograph; subsection~\ref{subsect:ex-eq-ex} gives a few examples of non-trivial relations meeting the requirements of the sufficient condition; subsection~\ref{subsect:a-eq-ex} applies the result to network routing.

\subsection{The Proof}\label{subsect:p-eq-ex}

The two main stages of the proof are, first, building hereditary maximal paths, which are the only paths involved in equilibria, and second, proceeding by induction on the number of arcs in the dalograph.

Hereditary maximal paths seem difficult to be built a priori. A weaker notion is that of semi-hereditary maximal path. On the one hand, a subpath of a hereditary maximal path is a maximal continuation of the preceding node along the hereditary maximal path. On the other hand, a subpath of a semi-hereditary maximal path is a maximal continuation of the beginning of the semi-hereditary maximal path, as defined below.

\begin{defn}[Semi-hereditary maximal path]
Let $x$ be a non-empty walk of continuation $\Gamma$. If $m(xy,\Gamma')$ for all decompositions $\Gamma=y\Gamma'$ where $xy$ is a walk, one writes $shm(x,\Gamma)$.
\end{defn}

Unsurprisingly, semi-hereditary maximality implies maximality.

\begin{lem}\label{lem:shm-m}
$shm(x,\Gamma)\,\Rightarrow\,m(x,\Gamma)$
\end{lem}

\begin{proof}
Instantiate $y$ with the empty walk in the definition of $shm$.
\end{proof}

The next result states that semi-hereditary maximality is implied by maximality plus semi-hereditary maximality of the subpath starting one node further along the path. 

\begin{lem}\label{lem:shm-m-shm}
$m(x,o\Gamma)\,\Rightarrow\,shm(xo,\Gamma)\,\Rightarrow\,shm(x,o\Gamma)$
\end{lem}

\begin{proof}
Consider a decomposition $o\Gamma=y\Gamma'$ where $xy$ is a walk. If $y$ is empty then $\Gamma'=o\Gamma$ so $m(xy,\Gamma')$ by assumption. If $y$ is not empty then $y=oz$. Since $\Gamma=z\Gamma'$ and $shm(xo,\Gamma)$,  it follows that $m(xoz,\Gamma')$. Hence $shm(x,o\Gamma)$.
\end{proof}

The notion of semi-hereditary maximality can also be defined along another induction principle for walks. In such a case, the definition would look like lemma~\ref{lem:shm-m-shm}.

As to the nibbling induction principle for walks in subsection~\ref{subsect:w-p}, it will be exploited as a recursive programming principle instead. It is used below to define a function that expects a non-empty walk and returns a path. It starts from one given node, finds a direction that promises maximality, follows the direction until there is a better direction to be followed, and so on, but without ever going back. It stops when the walk is looping because a looping walk defines a path. It is called the seeking-forward function.

\begin{defn}[Seeking-forward function]
Let $\prec$ be an acyclic preference and let $g$ be a dalograph. Define a function that expects a non-empty walk in $g$ and a continuation of this walk. More specifically, $\sff(x,\Gamma)$ is recursively defined along the nibbling induction principle for walks.
\begin{itemize}
\item If $x$ is a looping walk of continuation $\Gamma$, let $\sff(x,\Gamma)\eqdef\Gamma$.
\item If $x$ is not a looping walk then case split as follows.
\begin{enumerate}
\item If $m(x,\Gamma)$ then $\sff(x,\Gamma)\eqdef o\sff(xo,\Gamma')$, where $\Gamma=o\Gamma'$.
\item If $\neg m(x,\Gamma)$ then $\sff(x,\Gamma)\eqdef o\sff(xo,\Gamma')$ for some $o\Gamma'$ such that $m(x,o\Gamma')$ and $x^l\Gamma\prec x^lo\Gamma'$.
\end{enumerate}
\end{itemize} 
\end{defn}

The following lemma states that whatever the point of view, a path processed by the seeking-forward function is somehow not worse than the original path. Before reading the lemma recall that given $u$ a non-empty finite sequence, $u^f$ (resp. $u^l$) represents the first (resp. last) element of $u$.

\begin{lem}\label{lem:f-max}
Let $\prec$ be an irreflexive and A-transitive preference. Let $u$ be a non-empty suffix of $x$. The following formula holds.
\[u\sff(x,\Gamma)\prec u^f\Delta\,\Rightarrow\,u\Gamma\prec u^f\Delta\]
\end{lem}

\begin{proof}
By nibbling induction on walks. Base step, $x$ is a looping walk of unique continuation $\Gamma$. By definition of $\sff$ we have $\sff(x,\Gamma)=\Gamma$, so the claim holds. Inductive step, case split on $\Gamma$ being or not a maximal continuation of $x$. First case, $m(x,\Gamma)$, so $\sff(x,\Gamma)=o\sff(xo,\Gamma')$ with $\Gamma=o\Gamma'$. Assume that $u\sff(x,\Gamma)\prec u^f\Delta$, so $o\sff(xo,\Gamma')\prec u^f\Delta$. By induction hypothesis, $uo\Gamma'\prec u^f\Delta$, so $u\Gamma\prec u^f\Delta$. Second case, $\neg m(x,\Gamma)$, so $\sff(x,\Gamma)=o\sff(xo,\Gamma')$ and $x^l\Gamma\prec x^lo\Gamma'$ for some $o\Gamma'$ maximal continuation of $x$. Assume that $u\sff(x,\Gamma)\prec u^f\Delta$, so $uo\sff(xo,\Gamma')\prec u^f\Delta$. By induction hypothesis, $uo\Gamma'\prec u^f\Delta$. Since $u$ is a non-empty suffix of $x$, we have $u^l=x^l$. It follows that $u^l\Gamma\prec u^lo\Gamma'$. By A-transitivity, $u\Gamma\prec u^f\Delta$.
\end{proof}

The next lemma describes an involution property of the seeking-forward function, which suggests that seeking-forward once is enough.

\begin{lem}\label{lem:f-involutive}
Let $\prec$ be an irreflexive and A-transitive preference. The following formula holds.
\[\sff(x,\sff(x,\Gamma))=\sff(x,\Gamma)\]
\end{lem}

\begin{proof}
By nibbling induction on walks. Base step: $x$ is a looping walk of unique continuation $\Gamma$. By definition of $\sff$ we have $\sff(x,\Gamma)=\Gamma$, so $\sff(x,\sff(x,\Gamma))=\sff(x,\Gamma)$. Inductive step: by definition of $\sff$, $\sff(x,\Gamma)=o\sff(xo,\Gamma')$ for some $o\Gamma'$ such that $m(x,o\Gamma')$. So $\sff(x,\sff(x,\Gamma))=\sff(x,o\sff(xo,\Gamma'))$. Since $m(xo,\sff(xo,\Gamma'))$ by contraposition of lemma~\ref{lem:f-max}, $\sff(x,o\sff(xo,\Gamma'))=o\sff(xo,\sff(xo,\Gamma'))$. By induction hypothesis, $\sff(xo,\sff(xo,\Gamma'))=\sff(xo,\Gamma')$, so $\sff(x,\sff(x,\Gamma))=o\sff(xo,\Gamma')=\sff(x,\Gamma)$.
\end{proof}

The next analytical property about the seeking-forward function shows that a subpath of a fixed point is also a fixed point.

\begin{lem}\label{lem:sub-fp}
Let $\prec$ be an irreflexive and A-transitive preference, and let $xo$ be a walk. The following formula holds.
\[\sff(x,o\Gamma)=o\Gamma\,\Rightarrow\,\sff(xo,\Gamma)=\Gamma\]
\end{lem}

\begin{proof}
By definition of $\sff$, $\sff(x,o\Gamma)=o'\sff(xo',\Gamma')$ for some $o'\Gamma'$. It follows that $o'=o$ and $\sff(xo,\Gamma')=\Gamma$. Therefore $\sff(xo,\sff(xo,\Gamma'))=\sff(xo,\Gamma)$ by term substitution. By lemma~\ref{lem:f-involutive}, $\sff(xo,\sff(xo,\Gamma'))=\sff(xo,\Gamma')$, therefore $\sff(xo,\Gamma)=\Gamma$ by transitivity of equality.
\end{proof}

The following lemma states that any fixed point of the seeking-forward function is a semi-hereditary maximal path. The converse also holds and the proof is straightforward, but this converse result is not relevant in this chapter.

\begin{lem}\label{lem:fp-shm}
Let $\prec$ be an irreflexive, E-prefix, and A-transitive preference. The following formula holds.
\[\sff(x,\Gamma)=\Gamma\,\Rightarrow\,shm(x,\Gamma)\]
\end{lem}

\begin{proof}
By nibbling induction on the walk $x$. Base step, $x$ is a looping walk, so $m(x,\Gamma)$ by definition of $m$. For any decomposition $\Gamma=y\Gamma'$ where $xy$ is a walk, $y$ is empty and $\Gamma'=\Gamma$ because $x$ is a looping walk, so $m(xy,\Gamma')$. This shows $shm(x,\Gamma)$. Inductive step, assume that $\sff(x,o\Gamma)=o\Gamma$. So $\sff(xo,\Gamma)=\Gamma$ according to lemma~\ref{lem:sub-fp}, and $shm(xo,\Gamma)$ by induction hypothesis, so $m(xo,\Gamma)$ by lemma~\ref{lem:shm-m}. If $\neg m(x,o\Gamma)$, there exists a path $xo'\Gamma'$ such that $x^lo\Gamma\prec x^lo'\Gamma'$ and $\sff(x,o\Gamma)=o'\sff(xo',\Gamma')$. So $o'=o$, and $o\Gamma\prec o\Gamma'$ since $\prec$ is E-prefix, which contradicts $m(xo,\Gamma)$. Therefore $m(x,o\Gamma)$, and lemma~\ref{lem:shm-m-shm} allows concluding.
\end{proof}

The next lemma gives a sufficient condition so that semi-hereditary maximality implies hereditary maximality. 

\begin{lem}\label{lem:shm-hm}
Let $\prec$ be an irreflexive, E-prefix, and A-transitive preference whose inverse of negation is alt-subcontinuous, and let $g$ be a dalograph. 
\[shm_{g,\prec}(o,\Gamma)\,\Rightarrow\,hm_{g,\prec}( o\Gamma)\] 
\end{lem}

\begin{proof}
Assume that $shm(o,\Gamma)$. It suffices to prove by induction on the walk $x$ that $o\Gamma=xo_1\Gamma_1$ implies $m(o_1,\Gamma_1)$. First case, $x$ is empty, so $o=o_1$ and $\Gamma_1=\Gamma$. Since $shm(o,\Gamma)$ by assumption, $m(o_1,\Gamma_1)$ by definition of $shm$. Second case, assume that $o\Gamma=xo'o_1\Gamma_1$ and let $o_1\Gamma_2$ be a path. If $xo_1\Gamma_2$ is a path, $o\Gamma_1\not\prec o\Gamma_2$ by semi-hereditary maximality of $\Gamma$. If $xo_1\Gamma_2$ is not a path, then $x$ and $\Gamma_2$ intersect. Let $o_2$ be the first intersection node along $\Gamma_2$. So $x=uo_2v$ and $\Gamma_2=to_2\Gamma_3$, with $xo_1to_2$ being a looping walk. The situation is displayed below.

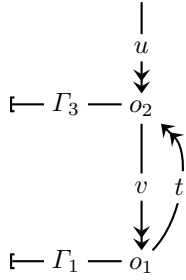
\[\psmatrix
  &&\phantom{aaaaaaaa}&[name=n]\\\\\\
  &[name=n1]&&[name=n2]o_2\\\\\\\\
  &[name=n3]&&[name=n4]o_1\\
  \ncline[nodesep=.1]{->>}{n}{n2}
  \mput*{u}
  \ncline[nodesep=.1]{-]}{n2}{n1}
  \mput*{\Gamma_3}
  \ncline[nodesep=.1]{->>}{n2}{n4}
  \mput*{v}
  \ncline[nodesep=.1]{-]}{n4}{n3}
  \mput*{\Gamma_1}
  \ncarc[nodesep=.1, arcangleA=-45,arcangleB=-45]{->>}{n4}{n2}
  \mput*{t}
\endpsmatrix\]

$o_1\Gamma_2$ is a path so $o_2\Gamma_3$ is also a path. Since $o\Gamma=uo_2vo_1\Gamma_1$ and $u$ is smaller than $x$, the induction hypothesis says that  $m(o_2,vo_1\Gamma_1)$. Therefore $o_2vo_1\Gamma_1\not\prec o_2\Gamma_3$. Because $xo_1to_2$ is a looping walk, $u(o_2vo_1t)^\omega$ is a path, by definition, and so is $(o_2vo_1t)^\omega$. Since $m(o_2,vo_1\Gamma_1)$, we have $ (o_2vo_1t)^\omega\prec o_2vo_1\Gamma_1$. Let $\alpha$ be the sequence induced by $o_1\Gamma_1$, $\beta$ by $o_2\Gamma_3$, $y$ by $o_2vo_1$, and $z$ by $o_1to_2$. We have $y\alpha\not\prec \beta$ and $\alpha\not\prec (zy)^\omega$, so $\alpha\not\prec z\beta$ by alt-subcontinuity of the inverse of the negation. Therefore $o_1\Gamma_1\not\prec o_1to_2\Gamma_3$, which shows that $m(o_1,\Gamma_1)$.
\end{proof}

Now that hereditary maximal paths are available/computable, it is possible to prove the existence of equilibrium for every dalograph, by induction on the number of arcs in the dalograph. Compute one hereditary maximal path and remove the arcs that the path ignored while visiting adjacent nodes. Compute an equilibrium on this smaller dalograph and add the ignored arcs back. This yields an equilibrium for the bigger dalograph. This procedure is detailed below.

\begin{thm}\label{thm:ESkr-eq}
If a preference is included in a subcontinuous E-prefix strict weak order, then any dalograph has a global equilibrium with respect to the preference. 
\end{thm}

\begin{proof}
According to lemma~\ref{lem:dalo-esp}, it suffices to show the claim for preferences that are actually subcontinuous E-prefix strict weak orders. Proceed by induction on the number of arcs in the dalograph. If the dalograph has one arc only, we are done since there is only one possible strategy and since preference is irreflexive. Now assume that the claim is proved for any dalograph with $n$ or less arcs, and consider a dalograph $g$ with $n+1$ arcs. If each node has only one outgoing arc, we are done by lemma~\ref{lem:no-choice-eq}. Now assume that there exists a node $o$ with at least two outgoing arcs as shown below.

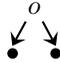
\[\psmatrix\\
  &&[name=n']o\\
  &[name=n1']\qdisk(0,0){2pt} &\phantom{aa}&[name=n2']\qdisk(0,0){2pt}\\
  \ncline[nodesep=.1]{->}{n'}{n1'}
  \ncline[nodesep=.1]{->}{n'}{n2'}
\endpsmatrix\]

Since the preference is a subcontinuous E-prefix strict weak order, it is irreflexive, A-transitive, E-prefix, and alt-subcontinuous by lemmas~\ref{lem:a-i},~\ref{lem:Ekr-ut} and~\ref{lem:aESc}. According to lemmas~\ref{lem:f-involutive},~\ref{lem:fp-shm} and~\ref{lem:shm-hm} there exists a hereditary maximal path $\Gamma$ starting from the node $o$. Let $u(av)^\omega$ be the corresponding sequences of labels, where $u$ and $v$ are finite (possibly empty) sequences of labels and $a$ is a label. 

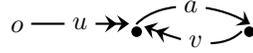
\[\psmatrix\\
  &[name=n1]o&\phantom{aaaaaaa}&[name=n2]\qdisk(0,0){2pt}&\phantom{aaaaaaa}&[name=n3]\qdisk(0,0){2pt}\\
  \ncline[nodesep=.1]{->>}{n1}{n2}
  \mput*{u}
  \ncarc[nodesep=.1, arcangleA=45,arcangleB=30]{->}{n2}{n3}
  \mput*{a}
  \ncarc[nodesep=.1, arcangleA=45,arcangleB=20]{->>}{n3}{n2}
  \mput*{v}
\endpsmatrix\]

From the dalograph $g$ and the path $\Gamma$, build a new dalograph $g'$ as follows. Remove all the arcs that are dismissed by the choices along $\Gamma$. There is at least one such arc since $\Gamma$ starts at node $o$, which has two or more outgoing arcs. Below, the dalograph $g$ is to the left and the dalograph $g'$ to the right. 

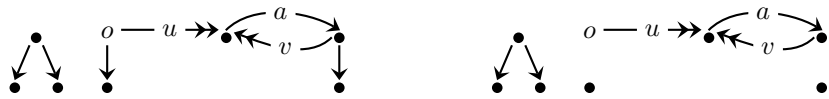
\[\begin{array}{c@{\hspace{2.0cm}}c}
\psmatrix\\
   &&[name=n']\qdisk(0,0){2pt}&&\phantom{aa}&[name=n1]o&\phantom{aaaaaaa}&[name=n2]\qdisk(0,0){2pt}&\phantom{aaaaaaa}&[name=n3]\qdisk(0,0){2pt}\\\\\\
   &[name=n1']\qdisk(0,0){2pt} &\phantom{aa}&[name=n2']\qdisk(0,0){2pt}&&[name=n4]\qdisk(0,0){2pt}&&&&[name=n5]\qdisk(0,0){2pt}\\
  \ncline[nodesep=.1]{->>}{n1}{n2}
  \mput*{u}
  \ncarc[nodesep=.1, arcangleA=45,arcangleB=30]{->}{n2}{n3}
  \mput*{a}
  \ncarc[nodesep=.1, arcangleA=45,arcangleB=20]{->>}{n3}{n2}
  \mput*{v}
  \ncline[nodesep=.1]{->}{n1}{n4}
  \ncline[nodesep=.1]{->}{n3}{n5}
  \ncline[nodesep=.1]{->}{n'}{n1'}
  \ncline[nodesep=.1]{->}{n'}{n2'}
\endpsmatrix
&
\psmatrix\\
   &&[name=n']\qdisk(0,0){2pt}&&\phantom{aa}&[name=n1]o&\phantom{aaaaaaa}&[name=n2]\qdisk(0,0){2pt}&\phantom{aaaaaaa}&[name=n3]\qdisk(0,0){2pt}\\\\\\
   &[name=n1']\qdisk(0,0){2pt} &\phantom{aa}&[name=n2']\qdisk(0,0){2pt}&&[name=n4]\qdisk(0,0){2pt}&&&&[name=n5]\qdisk(0,0){2pt}\\
  \ncline[nodesep=.1]{->>}{n1}{n2}
  \mput*{u}
  \ncarc[nodesep=.1, arcangleA=45,arcangleB=30]{->}{n2}{n3}
  \mput*{a}
  \ncarc[nodesep=.1, arcangleA=45,arcangleB=20]{->>}{n3}{n2}
  \mput*{v}
  \ncline[nodesep=.1]{->}{n'}{n1'}
  \ncline[nodesep=.1]{->}{n'}{n2'}
\endpsmatrix
\end{array}\]

 The new dalograph $g'$ has $n$ or less arcs, so it has an equilibrium by induction hypothesis. Such an equilibrium is represented below and named $s'$. The double lines represent the choices of the strategy, one choice per node.

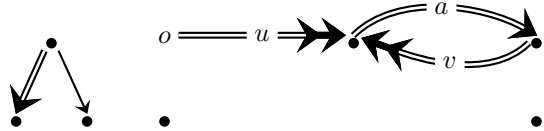
\[\psmatrix\\
   &&[name=n']\qdisk(0,0){2pt}&&\phantom{aaaa}&[name=n1]o&\phantom{aaaaaaaaaaaa}&[name=n2]\qdisk(0,0){2pt}&\phantom{aaaaaaaaaaaa}&[name=n3]\qdisk(0,0){2pt}\\\\
   &[name=n1']\qdisk(0,0){2pt} &\phantom{aaaa}&[name=n2']\qdisk(0,0){2pt}&&[name=n4]\qdisk(0,0){2pt}&&&&[name=n5]\qdisk(0,0){2pt}\\
  \ncline[nodesep=.1,doubleline=true]{->>}{n1}{n2}
  \mput*{u}
  \ncarc[nodesep=.1, arcangleA=45,arcangleB=30,doubleline=true]{->}{n2}{n3}
  \mput*{a}
  \ncarc[nodesep=.1, arcangleA=45,arcangleB=20,doubleline=true]{->>}{n3}{n2}
  \mput*{v}
  \ncline[nodesep=.1,doubleline=true]{->}{n'}{n1'}
  \ncline[nodesep=.1]{->}{n'}{n2'}
\endpsmatrix\]

The path induced by $s'$ from the node $o$ is the same as $\Gamma$ because there is only one possible path from this node in the dalograph $g'$, after removal of the arcs. From the equilibrium $s'$, build a new strategy named $s$ by adding back the arcs that were removed when defining $g'$, as shown below.

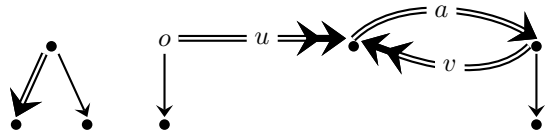
\[\psmatrix\\
   &&[name=n']\qdisk(0,0){2pt}&&\phantom{aaaa}&[name=n1]o&\phantom{aaaaaaaaaaaa}&[name=n2]\qdisk(0,0){2pt}&\phantom{aaaaaaaaaaaa}&[name=n3]\qdisk(0,0){2pt}\\\\
   &[name=n1']\qdisk(0,0){2pt} &\phantom{aaaa}&[name=n2']\qdisk(0,0){2pt}&&[name=n4]\qdisk(0,0){2pt}&&&&[name=n5]\qdisk(0,0){2pt}\\
  \ncline[nodesep=.1,doubleline=true]{->>}{n1}{n2}
  \mput*{u}
  \ncarc[nodesep=.1, arcangleA=45,arcangleB=30,doubleline=true]{->}{n2}{n3}
  \mput*{a}
  \ncarc[nodesep=.1, arcangleA=45,arcangleB=20,doubleline=true]{->>}{n3}{n2}
  \mput*{v}
  \ncline[nodesep=.1,doubleline=true]{->}{n'}{n1'}
  \ncline[nodesep=.1]{->}{n'}{n2'}
  \ncline[nodesep=.1]{->}{n1}{n4}
  \ncline[nodesep=.1]{->}{n3}{n5}
\endpsmatrix\]

The remainder of the proof shows that $s$ is an equilibrium for $g$. Since $\Gamma$ is hereditary maximal, $s$ is a local equilibrium for any node involved in $\Gamma$. Now let $o'$ be a node outside $\Gamma$, and let $\Gamma'$ be the path induced by $s$ (and $s'$) starting from $o'$. Consider another path starting from $o'$. If the new path does not involve any arc dismissed by $\Gamma$, then the new path is also valid in $g'$, so it is not greater than $\Gamma'$ since $s'$ is an equilibrium. If the new path involves such an arc, then the situation looks like the picture below, where $\Gamma=v\Gamma''$ and the new path is $u\Delta$.

\[\psmatrix\\
  &&&[name=n3]o'&&[name=n1]o\\\\\\
  &[name=n2]&\phantom{aaaaaaaaaa}&&\phantom{aaaaaaaa}&[name=n4]\qdisk(0,0){2pt}&\phantom{aaaaaaaaaa}&\\\\\\
  &&&&&[name=n6]&&[name=n5]\\
  \ncline[nodesep=.1]{-]}{n3}{n2}
  \mput*{\Gamma'}
  \ncline[nodesep=.1]{->>}{n3}{n4}
  \mput*{u}
  \ncline[nodesep=.1]{->>}{n1}{n4}
  \mput*{v}
  \ncline{-]}{n4}{n5} 
   \mput*{\Delta}
  \ncline[nodesep=.1]{-]}{n4}{n6}
   \mput*{\Gamma''}
  \endpsmatrix\]

By hereditary maximality, $\Gamma''\not\prec\Delta$, so $u\Gamma''\not\prec u\Delta$ by contraposition of E-prefix. The path $u\Gamma''$ is also valid in $g'$, so $\Gamma'\not\prec u\Gamma''$ because $s'$ is an equilibrium. Therefore $\Gamma'\not\prec u\Delta$ by transitivity of the negation. Hence, $\Gamma'$ is also maximal in $g$, and $s$ is an equilibrium.
\end{proof}

\subsection{Examples}\label{subsect:ex-eq-ex}

This subsection gives a few examples of non-trivial relations included in some subcontinuous E-prefix strict weak order. First, it discusses the lexicographic extension of a strict weak order, and a component-wise order in Pareto style, which happens to be included in the lexicographic extension. Second, it defines two limit-set orders involving maxima and minima of a set. The second order happens to be included in the first one.

\subsubsection{Lexicographic extension}

The lexicographic extension is widely studied in the literature. It is the abstraction of the way entries are ordered in a dictionary, hence the name. The lexicographic extension usually involves total orders, but it can be extended to strict weak orders.

\begin{defn}[Lexicographic extension]
Let $(A,\prec)$ be a set equipped with a strict weak order. The lexicographic extension of the strict weak order is defined over infinite sequences of elements of $A$.
\[\begin{array}{c@{\hspace{2cm}}c}
   \prooftree
    a\prec b
    \justifies
    a\alpha\prec^{lex}b\beta 
  \endprooftree
  &
 \prooftree
    \alpha\prec^{lex}\beta\phantom{aaaa} a\sharp b 
    \justifies
    a\alpha\prec^{lex}b\beta 
  \endprooftree
\end{array}\] 
\end{defn}

For instance $01^\omega\prec^{lex}10^\omega$ and $(03)^\omega\prec^{lex}1^\omega\prec^{lex}(30)^\omega$, with the usual order on figures $0$, $1$ and $3$.

The following defines when two sequences of the same length are "equivalent" with respect to a strict weak order.

\begin{defn}
Let $(A,\prec)$ be a set equipped with a strict weak order. Let $u$ and $v$ be two sequences of length $n$ of elements of $A$. If for all $i$ between $1$ and $n$, $u_i\sharp v_i$, then one writes $u\sharp v$.
\end{defn}

Next lemma characterises the lexicographic extension of a strict weak order through equivalent prefixes followed by comparable letters.

\begin{lem}\label{lem:lex-charact}
Let $(A,\prec)$ be a set equipped with a strict weak order, and let $\prec^{lex}$ be its lexicographic extension.
\[\begin{array}{c}
\alpha\prec^{lex}\beta\\
\Updownarrow\\
\exists u,v\in A^*,\exists a,b\in A,\exists\alpha',\beta'\in A^\omega,\,\alpha=ua\alpha'\,\wedge\,\beta=vb\beta'\,\wedge\, u\sharp v\,\wedge\,a\prec b
\end{array}\]
\end{lem}

\begin{proof}
left-to-right: by rule induction. First rule, $\alpha\prec^{lex}\beta$ comes from $a\prec b$, so $\alpha=a \alpha'$ and $\beta=b\beta'$ for some $\alpha'$ and $\beta'$. Second rule, $\alpha\prec^{lex}\beta$ comes from $\alpha'\prec^{lex}\beta'$ and $a\sharp b$. The induction hypothesis provides some $u$, $v$; $au$ and $bv$ are witnesses of for the claim. Right-to-left. By induction on the length of $u$. If $u$ is empty then the claim corresponds to the first inference rule. For the inductive step, invoke the second inference rule.
\end{proof}

The following lemma states the transitivity of the lexicographic extension of a strict weak order.

\begin{lem}\label{lem:lex-trans}
Let $(A,\prec)$ be a set equipped with a strict weak order. Then $\prec^{lex}$ is transitive.
\end{lem}

\begin{proof}
Assume that $\alpha\prec^{lex}\beta$ and $\beta\prec^{lex}\gamma$. By lemma~\ref{lem:lex-charact}, this gives one decomposition $\alpha=ua\alpha'$ and $\beta=ub\beta'$ with $a\prec b$, and one decomposition $\beta=vb'\beta''$ and $\gamma=vc\gamma'$ with $b'\prec c$. Therefore $\beta=ub\beta'=vb'\beta''$. Case split along the following three mutually exlusive cases: first $u=v$, second $u$ is a proper prefix of $v$, and third $v$ is a proper prefix of $u$. If $u=v$ then $b=b'$ so $a\prec c$ by transitivity of $\prec$, so $\alpha\prec^{lex}\gamma$ by lemma~\ref{lem:lex-charact}. If $u$ is a proper prefix of $v$ then $u=vb'v'$, so $\alpha=vb'v'\alpha'$ and $\gamma=vc\gamma'$ with $b'\prec c$, therefore $\alpha\prec^{lex}\gamma$ by lemma~\ref{lem:lex-charact}. If $v$ is a proper prefix of $u$ then $v=ubu'$, so $\alpha=ua\alpha'$ and $\gamma=ubu'\gamma'$ with $a\prec b$, therefore $\alpha\prec^{lex}\gamma$ by lemma~\ref{lem:lex-charact}.
\end{proof}

By contraposition, lemma~\ref{lem:lex-charact} yields the following characterisation of $\not\prec^{lex}$, the negation of $\prec^{lex}$.

\begin{lem}\label{lem:notlex-charact}
Let $(A,\prec)$ be a set equipped with a strict weak order, and let $\prec^{lex}$ be its lexicographic extension.
\[\begin{array}{c}
\alpha\not\prec^{lex}\beta\\
\Updownarrow\\
\forall u,v\in A^*,\forall a,b\in A,\forall\alpha',\beta'\in A^\omega,\,\alpha=ua\alpha'\,\wedge\,\beta=vb\beta'\,\wedge\, u\sharp v\,\Rightarrow a\not\prec b
\end{array}\]
\end{lem}

The construction of the lexicographic extension preserves strict weak ordering, as stated below.

\begin{lem}\label{lem:lex-swo}
Let $(A,\prec)$ be a set equipped with a strict weak order. The derived $\prec^{lex}$ is also a strict weak order.
\end{lem}

\begin{proof}
Since $\prec^{lex}$ is transitive by lemma~\ref{lem:lex-trans}, it suffices to show that $\sharp^{lex}$ is an equivalence relation, by lemma~\ref{lem:kn-teq-ic}. The relation $\prec^{lex}$ is irreflexive, which can be proved by rule induction on its definition. So $\sharp^{lex}$ is reflexive. It is also symmetric by definition. By lemma~\ref{lem:notlex-charact}, $\alpha\sharp^{lex}\beta$ is equivalent to $u\sharp v$ for all decompositions $\alpha=u\alpha'$ and $\beta=v\beta'$ with $u$ and $v$ of the same length. This property is transitive since $\sharp$ is transitive by lemma~\ref{lem:kn-teq-ic}.
\end{proof}

The lexicographic extension of a strict weak order is E-prefix.

\begin{lem}\label{lem:lex-Epref}
Let $(A,\prec)$ be a set equipped with a strict weak order. The derived $\prec^{lex}$ is E-prefix.
\end{lem}

\begin{proof}
Prove by induction on $u$ that $u\alpha\prec^{lex}u\beta$ implies $\alpha\prec^{lex}\beta$. If $u$ is empty, that is trivial. If $u=au'$, then $au'\alpha\prec^{lex}au'\beta$ must come from the second inference rule of the definition of $\prec^{lex}$, which means that $u'\alpha\prec^{lex}u'\beta$. Therefore  $\alpha\prec^{lex}\beta$ by induction hypothesis.
\end{proof}

The lexicographic extension of a strict weak order is also subcontinuous.

\begin{lem}\label{lem:lex-cons}
Let $(A,\prec)$ be a set equipped with a strict weak order. The derived $\prec^{lex}$ is subcontinuous.
\end{lem}

\begin{proof}
Assume that $u^\omega\prec^{lex}\alpha$, so $u^nu'\sharp v$, $u=u'au''$, $\alpha=vb\alpha'$, and $a\prec b$ for some $n$, $u'$, $u''$, $a$, $b$ and $\alpha'$. Therefore $u^{n+1}u'\sharp uv$, which can be written $u^nu'au''u'\sharp uv$. Decompose $uv=v'cv''$ with $v''$ and $u''u'$ of the same length. So $u^nu'\sharp v'$ and $a\sharp c$. So $c\prec b$ by strict weak ordering and $v\sharp v'$ since $\sharp$ is an equivalence relation. Since $u\alpha=v'c(v''b\alpha')$ and $\alpha=vb\alpha'$, $u\alpha\prec^{lex}\alpha$. Therefore $\prec^{lex}$ is subcontinuous.
\end{proof}

Theorem~\ref{thm:ESkr-eq} together with lemmas~\ref{lem:lex-swo},~\ref{lem:lex-Epref}, and~\ref{lem:lex-cons} allows stating the following.

\begin{thm}\label{thm:lex-eq}
A dalograph labelled with elements of a strict weak order has a global equilibrium with respect to the lexicographic extension of the strict weak order.
\end{thm}

\subsubsection{Pareto Extension}

A Pareto extension allows comparing vectors with comparable components. A first vector is "greater" than a second one if it is not smaller component-wise and if it is greater for some component. This can be extended to infinite sequences.

\begin{defn}[Pareto extension]
Let $(A,\prec)$ be a set equipped with a strict weak order. The Pareto extension of the strict weak order is defined over infinite sequences of elements of $A$.
\begin{eqnarray*}
\alpha\prec^P\beta &\quad\eqdef\quad& \forall n\in\mathbb{N}, \beta(n)\not\prec\alpha(n)\quad\wedge\quad\exists n\in\mathbb{N},\, \alpha(n)\prec\beta(n)
\end{eqnarray*}
\end{defn}

For instance $01^\omega\not\prec^{P}10^\omega$ but $(01)^\omega\prec^{P}1^\omega\prec^{P}(13)^\omega$ with the usual order on figures $0$, $1$ and $3$.

The following lemma states that a Pareto extension of a strict weak order is included in the lexicographic extension of the same strict weak order.

\begin{lem}\label{lem:pareto-lex}
Let $(A,\prec)$ be a set equipped with a strict weak order.
\[\alpha\prec^P\beta\quad\Rightarrow\quad\alpha\prec^{lex}\beta\]
\end{lem}

\begin{proof}
Assume that $\alpha\prec^P\beta$, so by definition $\beta(n)\not\prec\alpha(n)$ for all naturals $n$, and $\alpha(n)\prec\beta(n)$ for some natural $n$. Let $n_0$ be the smallest natural $n$ such that $\alpha(n)\prec\beta(n)$. So $\alpha=ua\alpha'$ and $\beta=vb\alpha$ for some $u$ and $v$ of length $n_0$ and $a\prec b$. For $i$ between $0$ and $n_0-1$, $v(i)\not\prec u(i)$ by assumption, and $u(i)\not\prec v(i)$ by definition of $n_0$. Therefore $u\sharp v$, so  $\alpha\prec^{lex}\beta$ by lemma~\ref{lem:kn-teq-ic}.
\end{proof}

Pareto extension also guarantees existence of equilibrium in dalographs.

\begin{thm}
A dalograph labelled with elements of a strict weak order has a global equilibrium with respect to the derived Pareto extension.
\end{thm}

\begin{proof}
Invoke lemmas~\ref{lem:pareto-lex} and~\ref{lem:dalo-esp}, and theorem~\ref{thm:lex-eq}.
\end{proof}

\subsubsection{Max-Min Limit-Set Order}

As discussed in subsection~\ref{subsect:gbr}, two non-comparable elements of a strict weak order compare the same way against any third element. Therefore, comparison of two elements amounts to comparison of their non-comparability equivalence classes.

\begin{defn}
Let $(E,\prec)$ be a set equipped with a strict weak order. This induces a total order defined as follows on the $\sharp$-equivalence classes $A^\sharp$ and $B^\sharp$.
\[A^\sharp\prec B^\sharp\quad\eqdef\quad\exists x\in A^\sharp,\exists y\in B^\sharp,\,x\prec y\]
\end {defn}

Through total ordering, it is easy to define a notion of extrema of finite sets.

\begin{defn}[Class maximum and minimum]
Let $(E,\prec)$ be a set equipped with a strict weak order. The maximum (resp. minimum) of a finite subset $A$ of $E$ is the maximal (resp. minimal) $\sharp$-class intersecting $A$.
\end{defn}

An order over sets is defined below. It involves extrema of sets.

\begin{defn}[Max-min order over sets]
Let $(E,\prec)$ be a set equipped with a strict weak order. The max-min order is defined on finite subsets of $E$.
\[A\prec^{Mm}B\quad\eqdef\quad max(A)\prec max(B)\,\vee\,(max(A)=max(B)\,\wedge\,min(A)\prec min(B))\]
\end{defn}

For instance $\{1,2,3\}\prec^{Mm}\{0,4\}$ and $\{0,2,3\}\prec^{Mm}\{1,3\}$ with the usual total order over the naturals.

The negation of the above order is characterised below.

\begin{lem}\label{lem:not-prec-Mm}
Let $(E,\prec)$ be a set equipped with a strict weak order.
\[A\not\prec^{Mm}B\,\Leftrightarrow\, max(A)\not\prec max(B)\,\wedge\,(max(A)=max(B)\,\Rightarrow\,min(A)\not\prec min(B))\]
\end{lem}

The max-min construction preserves strict weak ordering, as stated below.

\begin{lem}
Let $(E,\prec)$ be a set equipped with a strict weak order. The Max-min order on $E$ is also a strict weak order.
\end{lem}

\begin{proof}
First, prove transitivity of its negation. Assume that $A\not\prec^{Mm}B$ and $B\not\prec^{Mm}C$. By assumption, $max(A)\not\prec max(B)$ and $max(B)\not\prec max(C)$, so $max(A)\not\prec max(C)$ since $\prec$ is a total order for $\sharp$-classes. Assume that $max(A)=max(C)$, so $max(A)=max(C)=max(B)$. Therefore $min(A)\not\prec min(B)$ and $min(B)\not\prec min(C)$ follows from the assumptions, and $min(A)\not\prec min(C)$ by total ordering. This shows that $A\not\prec^{Mm}C$. Second, prove that $\sharp^{Mm}$ is an equivalence relation: just note that $A\sharp^{Mm}B$ is equivalent to $max(A)=max(B)$ and $min(A)=min(B)$.
\end{proof}

The elements that appear infinitely many times in an infinite sequence constitute the limit set of the sequence. Sequences with non-empty limit sets can be compared through max-min comparisons of their limit sets.

\begin{defn}[Max-min limit-set order] Let $(E,\prec)$ be a set equipped with a strict weak order. For $\alpha$ infinite sequence over $E$, let $L_\alpha$ be its limit set, \textit{i.e.} the set of the elements that occur infinitely often in $\alpha$. Two infinite sequences whose limit sets are non-empty are compared as follows.
\[\alpha\prec^{Mmls}\beta\quad\eqdef\quad L_\alpha\prec^{Mm}\L_\beta\]
\end{defn}

For instance $3^n4^\omega\prec^{Mmls}0^p(50^q)^\omega$ because $4\prec 5$ according to the usual order over the naturals.

Next lemma states preservation of strict weak ordering by the max-min construction.

\begin{lem}\label{lem:Mm-ls-swo}
Let $(E,\prec)$ be a set equipped with a strict weak order. The max-min limit-set order is a strict weak order over the sequences with non-empty limit set.
\end{lem}

Since the limit set of a sequence is preserved by prefix elimination and addition, the following holds.

\begin{lem}\label{lem:Mm-ls-Epref-cons}
Let $(E,\prec)$ be a set equipped with a strict weak order. The max-min limit-set order is E-prefix and subcontinuous over the sequences of non-empty limit set.
\end{lem}

Theorem~\ref{thm:ESkr-eq} together with lemmas~\ref{lem:Mm-ls-swo} and~\ref{lem:Mm-ls-Epref-cons} allows stating the following.

\begin{thm}\label{thm:Mm-ls-eq}
A dalograph labelled with elements of a strict weak order has a global equilibrium with respect to the derived max-min limit-set order.
\end{thm}

\subsubsection{Max-Min-Light Limit-Set Order}

Roughly speaking, the max-min-light order relates sets such that the elements of one are bigger than the elements of the other.

\begin{defn}[Max-min-light order]
Let $(E,\prec)$ be a set equipped with a strict weak order. The max-min-light order is defined on finite subsets of $E$.
\[A\prec^{Mml}B\quad\eqdef\quad\forall x\in A,\forall y\in B,\,y\not\prec x\quad\wedge\quad\exists x\in A,\exists y\in B,\, x\prec y\]
\end{defn}

For instance $\{1,2\}\not\prec^{Mml}\{0,3\}$ but $\{0,1\}\prec^{Mml}\{1\}\prec^{Mml}\{1,2\}$

\begin{defn}[Max-min-light limit-set order] Let $(E,\prec)$ be a set equipped with a strict weak order. For $\alpha$ infinite sequence over $E$, let $L_\alpha$ be its limit set, \textit{i.e.} the set of the elements that occur infinitely often in $\alpha$. Two infinite sequences whose limit sets are non-empty are compared as follows.
\[\alpha\prec^{Mmlls}\beta\quad\eqdef\quad L_\alpha\prec^{Mml}\L_\beta\]
\end{defn}

The following theorem states that the max-min-light limit-set order guarantees equilibrium existence.

\begin{thm}
A dalograph labelled with elements of a strict weak order has a global equilibrium with respect to the derived max-min-light limit-set order.
\end{thm}

\begin{proof}
Note that the max-min-light limit-set order is included in the max-min limit-set order. Conclude by lemma~\ref{lem:dalo-esp} and theorem~\ref{thm:Mm-ls-eq}.
\end{proof}

\subsection{Application to Network Routing}\label{subsect:a-eq-ex}

The following issue is related to existing literature such as~\cite{GS05}.

\begin{defn}
A \emph{routing policy} is a binary relation over finite words over a collection of labels. A \emph{routing problem} is a finite digraph whose arcs are labelled with the above-mentioned labels. In addition, one node is called the \emph{target}. It has outdegree zero and it is reachable from any node through some walk in the digraph. A \emph{routing strategy} for the routing problem is a function mapping every node different from the target to one of the arcs going out that node. A routing strategy is said to be a \emph{routing equilibrium} if for each node, the path that is induced by the strategy starting from that node leads to the target, and if for each node, no strategy induces a better (according to the routing policy) such path. 
\end{defn}

The following lemma gives a sufficient condition for every routing problem to have a routing equilibrium. The condition may not be decidable in general, but it is decidable on the domain of every finite routing problem.

\begin{lem}\label{lem:routing-eq-impl}
If a routing policy is (included in) an E-prefix strict weak order $\prec^r$ such that $v\not\prec^ruv$ for all $v$ and $u$, then every routing problem has a routing equilibrium.
\end{lem}

\begin{proof}
A routing problem can be transformed into a dalograph as follows. Add a dummy node to the routing problem, which is a digraph. Add an arc from the target to the dummy node and from the dummy node to itself. Add dummy labels $dl$ on both arcs. From the routing policy $\prec^r$ over finite words, build a preference $\prec$ over the union of two sets. The first set is made of the infinite words over the original labels $L$ (without the dummy label). The second set is made of the concatenations of finite words over the original labels and the infinite word $dl^\omega$ built only with the dummy label.
\[\begin{array}{c@{\hspace{2cm}}c}
   \prooftree
    u\prec^r v
    \justifies
    udl^\omega\prec vdl^\omega
  \endprooftree
  &
 \prooftree
    \alpha\in L^\omega\phantom{aaa}u\in L^* 
    \justifies
    \alpha\prec udl^\omega 
  \endprooftree
\end{array}\] 
Since $\prec^r$ is E-prefix by assumption, $\prec$ is also E-prefix. Since $v\not\prec^ruv$ for all $v$ and $u$ by assumption, $\prec$ is subcontinuous. Therefore the built dalograph has a global equilibrium, by theorem~\ref{thm:ESkr-eq}. This global equilibrium corresponds to a routing equilibrium.
\end{proof}

\section{Simple closures}\label{sect:pc}

This section introduces the notion of simple closure, which characterises some operators on relations, and the notion of union of simple closures. This yields two monoids whose combination is similar to a semiring (distributivity is in the opposite direction though). This development intends to show that if given simple closures preserve a predicate, then any finite restriction of their union also preserves this predicate. This result will be usefull in section~\ref{sect:p-eqex}. 

The section starts with the following general lemma. It states that if a predicate is preserved by given functions, then it is preserved by any composition of these functions.

\begin{lem}\label{lem:comp-pre}
Let $f_0$ to $f_n$ be functions of type $A\to A$. Let $Q$ be a predicate on~$A$. Assume that each $f_k$ preserves $Q$. Then for all $x$ in $A$ and all $w$ words on the $f_k$ the following formula holds.
\[Q(x)\,\Rightarrow\,Q(w(x))\]
\end{lem}

\begin{proof}
By induction on $w$. First case, $w$ is the empty word, \textit{i.e.} the identity function. So $x$ equals $w(x)$, and $Q(x)$ implies $Q(w(x))$. Second case, $w$ equals $f_kw'$ and the claim holds for $w'$. Assume $Q(x)$, so $Q(w'(x))$ by induction hypothesis, and $Q(f_k\circ w'(x))$ since $f_k$ preserves $Q$.
\end{proof}

In the remainder of this section, the function domain $A$ that is mentioned in lemma~\ref{lem:comp-pre} above will be a set of relations. More specifically, $A$ will be the relations of arity $r$, for an arbitrary $r$ that is fixed throughout the section. In addition, the symbol $X$ represents a vector $(X_1,\dots,X_r)$ of dimension $r$.

Usually in mathematics, the closure of an object is the smallest bigger (or equal) object of same type that complies with some given predicates. What follows describes a certain kind of closures for relations of arity $r$, namely simple closures. Simple closures are operators (on relations of araity $r$) inductively defined through inference rules. The first rule ensures that the simple closures are bigger or equal than the original relation. The other rules pertain to the intended properties of the simple closure.

\begin{defn}
Let $f$ be an operator on relations of arity $r$. The operator $f$ is said to be a simple closure if it is inductively defined with the first inference rule below and some rules having the same form as the second inference rule below.
\[\prooftree
     R(X)
    \justifies
    f(R)(X)
  \endprooftree\]
\[\prooftree
    K(X, \{X^i\}_{i\in C}) \qquad \wedge_{i\in C} f(R)(X^i)
    \justifies
    f(R)(X)
  \endprooftree\]
\end{defn}

The next lemma states a few basic properties involving simple closures and inclusion.

\begin{lem}\label{lem:clos-sub}
Let $f$ be a simple closure. The following formulae hold.
\begin{itemize}
\item $R\subseteq f(R)$
\item $R\subseteq R'\,\Rightarrow\,f(R)\subseteq f(R')$
\item $f\circ f(R)\subseteq f(R)$
\end{itemize}
\end{lem}

\begin{proof}
The first claim follows the first rule $R(X)\,\Rightarrow\, f(R)(X)$. The second claim is proved by rule induction on the definition of $f$. First case, $f(R)(X)$ comes from $R(X)$. By inclusion, $R'(X)$, so $f(R')(X)$. Second case $f(R)(X)$ comes from $\wedge_{i\in C} f(R)(X^i)$. By induction hypothesis, $\wedge_{i\in C} f(R')(X^i)$, therefore $f(R')(X)$. The third claim is also proved by rule induction on the definition of $f$. First case, $f\circ f(R)(X)$ comes from $f(R)(X)$ we are done. Second case $f\circ f(R)(X)$ comes from $\wedge_{i\in C} f\circ f(R)(X^i)$. By induction hypothesis, $\wedge_{i\in C} f(R)(X^i)$, so $f(R)(X)$.
\end{proof}

The following lemma generalises the first property of lemma~\ref{lem:clos-sub}.

\begin{lem}\label{lem:clos-incr}
Let $f_0$ to $f_n$ be simple closures on relations of arity $r$, and Let $w$, $u$, and $v$ be words on the $f_k$. For all $R$ relation of arity $r$ and for all $X$ vector of dimension $r$, the following formula holds.
\[w(R)(X)\,\Rightarrow\,uwv(R)(X)\]
\end{lem}

\begin{proof}
First, prove the claim for empty $v$ by induction on $u$. If $u$ is empty then it is trivial. If $u=f_iu'$ then $u'w(R)(X)$ by induction hypothesis, so $f_iu'w(R)(X)$ by the first part of lemma~\ref{lem:clos-sub}. Second, prove the claim for empty $u$ by induction on $v$. If $v$ is empty then it is trivial. If $v=v'f_i$ then $wv'(R)(X)$ by induction hypothesis. Since $R\subseteq f_i(R)$ by the first part of lemma~\ref{lem:clos-sub}, we also have  $wv'f_i(R)(X)$ by the second part of lemma~\ref{lem:clos-sub}. Third, assume $w(R)(X)$. So $uw(R)(X)$ by the first part of this proof, and $uwv(R)(X)$ by the second part of this proof.
\end{proof}

\begin{defn}[Rule union]
Let $f$ and $g$ be two simple closure on relations of arity $r$. Assume that $f$ is defined by the induction rules $F_1$ to $F_n$, and that $g$ is defined by the induction rules $G_1$ to $G_m$. Then, the operator $f+g$ defined by the induction rules $F_1$ to $F_n$ and $G_1$ to $G_m$ is also a simple closure on relations of arity $r$. 
\end{defn}

The law $+$ defines an abelian monoid on simple closures on relation of the same arity, the neutral element being the identity operator. Moreover the law $+$ is distributive over the law $\circ$, but it should be the opposite for $(A\to A,+,\circ)$ to be a semiring. 

The union of two simple closures yields bigger relations than simple closures alone, as stated below. It is provable by rule induction on the definition of $f$.

\begin{lem}\label{lem:union-big}
Let $R$ be a relation and let $f$ and $g$ be simple closures. The following formula holds.
\[f(R)\subseteq (f+g)(R)\]
\end{lem}

The following lemma generalises the previous result. It shows that composition is somehow "bounded" by union. 

\begin{lem}\label{lem:union-comp}
Let $f_0$ to $f_n$ be simple closures on relations of arity $r$, and let $f$ equal $\Sigma_{0\leq k\leq n}f_k$. Let $w$ be a word on the $f_k$. For all $R$ relation of arity $r$, we have $w(R)\subseteq f(R)$.
\end{lem}

\begin{proof}
By induction on $w$. If $w$ is empty then $w(R)=R$ and the first part of lemma~\ref{lem:clos-sub} allows concluding. If $w=f_iw'$ then $w'(R)\subseteq f(R)$ by induction hypothesis. So $f_iw'(R)\subseteq f\circ f(R)$ by lemma~\ref{lem:union-big}, and $f_iw'(R)\subseteq f(R)$ by the third part of lemma~\ref{lem:clos-sub}.
\end{proof}

Although composition is "bounded" by union, union is approximable by composition, as developed in the next two lemmas. For any relation $R$ of arity $r$, the union of given simple closures (applied to $R$) can be simulated at a given point $X$ by some composition of the same simple closures (applied to $R$), as stated below.

\begin{lem}\label{lem:it-comp}
Let $f_0$ to $f_n$ be simple closures on relations of arity $r$, and let $f$ equal $\Sigma_{0\leq k\leq n}f_k$. Let $R$ be a relation of arity $r$. If $f(R)(X)$ then $w(R)(X)$ for some word $w$ on the $f_k$.
\end{lem}

\begin{proof}
By rule induction on the definition of $f$. First case, assume that $f(R)(X)$ comes from the following rule.
\[\prooftree
     R(X)
    \justifies
    f(R)(X)
  \endprooftree\]
Since $R(X)$ holds, $id(R)(X)$ also holds, so the empty word is a witness for the claim. Second case, assume that $f(R)(X)$ is induced by the following rule.
\[
   \prooftree
    K(X, \{X^i\}_{i\in C}) \qquad \wedge_{i\in C} f(R)(X^i)
    \justifies
    f(R)(X)
  \endprooftree
\]
By induction hypothesis, for all $i$ in $C$, $f(R)(X^i)$ implies that $w_i(R)(X^i)$ for some word $w_i$.
Let $w$ be a concatenation of the $w_i$. By lemma~\ref{lem:clos-incr}, we have $w(R)(X^i)$ for all $i$. Assume that the inference rule above comes from $f_k$. By lemma~\ref{lem:clos-incr}, $f_{k+1}\dots f_{n}w(R)(X^i)$ holds for all $i$. So, $f_kf_{k+1}\dots f_{n}w(R)(X)$ by applying the inference rule. So $f_{0}\dots f_{n}w(R)(X)$ by lemma~\ref{lem:clos-incr} again. Therefore the word $f_{0}\dots f_{n}w$ is a witness, whatever $f_k$ the inference rule may come from.
\end{proof}

For any relation $R$ of subdomain $S$, the union of given simple closures (applied to $R$) can be approximated on $S$ by a composition of the same simple closures (applied to $R$), as stated below.

\begin{lem}\label{lem:it-incl}
Let $f_0$ to $f_n$ be simple closures on relations of arity $r$, and let $f$ equal $\Sigma_kf_k$. Let $R$ be a relation of arity $r$, and let $S$ be a finite subdomain of $R$. There exists a word $w$ on the $f_k$ such that $f(R)\mid_S$ is included in $w(R)$. 
\end{lem}

\begin{proof}
Since $S$ is finite, there exist finitely many $X$  in $S$ such that $f(R)(X)$. For each such $X$ there exists a word $u$ such that $u(R)(X)$, by lemma~\ref{lem:it-comp}. Let $w$ be a concatenation of all these $u$. By lemma~\ref{lem:clos-incr}, $w(R)(X)$ for each such $X$. Therefore $f(R)\mid_S$ is included in $w(R)$.
\end{proof}

The following lemma shows that if given simple closures preserve a predicate that is also preserved by subrelation, then any finite restriction of the union of the closures also preserves the predicate.

\begin{lem}\label{lem:union-pres}
Let $Q$ be a predicate on relations that is preserved by the simple closures $f_0$ to $f_n$ and subrelation, \textit{i.e.} $R\subset R'\,\Rightarrow\,Q(R')\,\Rightarrow\,Q(R)$. Then for all relations $R$ of finite subdomain $S$, $Q(R)$ implies $Q(\Sigma_kf_k(R)\mid_S)$.
\end{lem}

\begin{proof}
Lemma~\ref{lem:it-incl} provides a $w$ such that $f(R)\mid_S$ is included in $w(R)$. Then lemma~\ref{lem:comp-pre} shows that $Q(w(R))$, and preservation by subrelation allows concluding.
\end{proof}

\section{Preservation of Equilibrium Existence}\label{sect:p-eqex}

This section defines (gen-) E-prefix and A-transitive closure. These are closely related to the (gen-) E-prefix and A-transitivity predicates that are defined in subsection~\ref{subsect:brs}. It is shown that these closures preserve equilibrium existence, \textit{i.e.} if every dalograph has an equilibrium with respect to a preference, then every dalograph also has a  equilibrium with respect to the closure of the preference. A combination of these closures is defined, and it also preserves equilibrium existence.

The E-prefix closure of a binary relation is its smallest E-prefix superrelation. It is inductively defined below.

\begin{defn}[E-prefix closure]
\[\begin{array}{c@{\hspace{.8cm}}c}
   \prooftree
    \alpha\prec\beta 
    \justifies
    \alpha\prec^{ep}\beta 
  \endprooftree
  &
  \prooftree
    u\alpha\prec^{ep}u\beta
    \justifies
    \alpha\prec^{ep}\beta 
  \endprooftree
\end{array}\] 
\end{defn}

The following lemma states that if a preference guarantees existence of equilibrium for all dalographs, then the E-prefix closure of this preference also  guarantees existence of equilibrium for all dalographs.

\begin{lem}\label{lem:epref-pres}
E-prefix closure preserves existence of equilibrium. Put otherwise, if all dalographs have $\prec$-equilibria, then all dalographs have $\prec^{ep}$-equilibria.
\end{lem}

\begin{proof}
Let $g$ be a dalograph. First note that, if $\alpha\prec^{ep}\beta$, then there exists $u$ such that $u\alpha\prec u\beta$. (Provable by rule induction). At any node with eligible $\alpha$ and $\beta$ such that $\alpha\prec^{ep}\beta$, add an incoming path inducing $u$, as shown below.

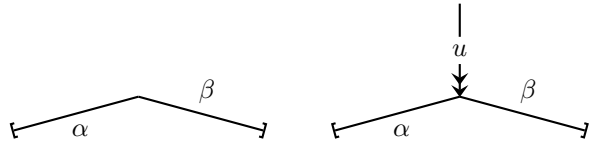
\[\begin{array}{c@{\hspace{.8cm}}c}
\psmatrix
  &&&[name=n]\\\\
  &[name=n1]&\phantom{aaaaaaaa}&&\phantom{aaaaaaaa}&[name=n2]\\
  \ncline[nodesep=0]{-]}{n}{n1}
  \Aput{\alpha}
  \ncline[nodesep=0]{-]}{n}{n2}
  \Aput{\beta}	
\endpsmatrix
&
\psmatrix
  &&&[name=n0]\\\\\\\\\\\\\\
  &&&[name=n]\\\\
  &[name=n1]&\phantom{aaaaaaaa}&&\phantom{aaaaaaaa}&[name=n2]\\
  \ncline[nodesep=.1]{->>}{n0}{n}
  \mput*{u}
  \ncline[nodesep=0]{-]}{n}{n1}
  \Aput{\alpha}
  \ncline[nodesep=0]{-]}{n}{n2}
  \Aput{\beta}	
\endpsmatrix
\end{array}\]

This new dalograph $g'$ has a $\prec$-equilibrium. For any $\alpha$ and $\beta$ such that $\alpha\prec^{ep}\beta$, the equilibrium does not induce $u\alpha$ when $u\beta$ is possible, so it does not induce $\alpha$ when $\beta$ is possible. Removing the newly added walks $u$ yields a $\prec^{ep}$-equilibrium.
\end{proof}

The transitive closure of a binary relation is its smallest transitive superrelation. It is inductively defined below according to the usual formal definition of transitive closure.

\begin{defn}[Transitive closure]
\[\begin{array}{c@{\hspace{.8cm}}c}
   \prooftree
    \alpha\prec\beta 
    \justifies
    \alpha\prec^{t}\beta 
  \endprooftree
  &
  \prooftree
    \alpha\prec^{t}\beta \qquad  \beta\prec^{t}\gamma
    \justifies
    \alpha\prec^{t}\gamma 
  \endprooftree
\end{array}\] 
\end{defn}

For the transitive closure and other closures that are dealt with in this chapter, it may not be as simple as for E-prefix closure to prove preservation of equilibrium existence. It is done in two steps in this chapter.

\begin{lem}\label{lem:trans-step}
Let $\prec$ be a preference, and let $\alpha$ and $\beta$ be such that $\alpha\prec^{t}\beta$. There exists a dalograph $g$ with the following properties.

\begin{itemize}
\item Only one "top" node has several outgoing arcs.
\item The dalograph below is a subgraph of the dalograph $g$. 

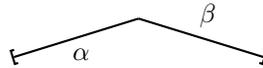
\[\psmatrix
  &&&[name=n]\\
  &[name=n1]&\phantom{aaaaaaaa}&&\phantom{aaaaaaaa}&[name=n2]\\
  \ncline[nodesep=0]{-]}{n}{n1}
  \Aput{\alpha}
  \ncline[nodesep=0]{-]}{n}{n2}
  \Aput{\beta}	
\endpsmatrix\]

\item Only $\beta$ may be $\prec$-maximal among the eligible sequences at the top node.
\end{itemize}
\end{lem}

\begin{proof}
Proceed by rule induction on the definition of the transitive closure. First rule, $\alpha\prec^{t}\beta$ comes from $\alpha\prec\beta$. The dalograph below complies with the requirements.

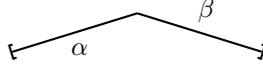
\[\psmatrix
  &&&[name=n]\\
  &[name=n1]&\phantom{aaaaaaaa}&&\phantom{aaaaaaaa}&[name=n2]\\
  \ncline[nodesep=0]{-]}{n}{n1}
  \Aput{\alpha}
  \ncline[nodesep=0]{-]}{n}{n2}
  \Aput{\beta}	
\endpsmatrix\]

Second rule, $\alpha\prec^{t}\beta$ comes from $\alpha\prec^{t}\gamma$ and $\gamma\prec^{t}\beta$. By induction hypothesis there exists one dalograph for $\alpha\prec^{t}\gamma$ and one for $\gamma\prec^{t}\beta$, as shown below to the left and the centre. In both dalographs, there is a node with several outgoing arcs. Fuse these nodes as shown below on the right-hand side.

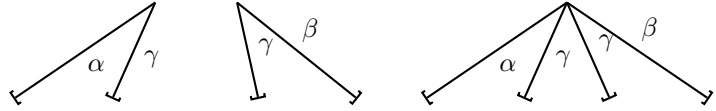
\[\begin{array}{l@{\hspace{.8cm}}l@{\hspace{.8cm}}r}
\psmatrix
  &&&&&[name=n]\\\\\\\\\\\\
  &[name=n1]&\phantom{aaaaaa}&[name=n2]&\phantom{aa}\\
  \ncline[nodesep=0]{-]}{n}{n1}
  \Aput{\alpha}
  \ncline[nodesep=0]{-]}{n}{n2}
  \Aput{\gamma}	
\endpsmatrix
&
\psmatrix
  &[name=n]\\\\\\\\\\\\
  &\phantom{aa}&[name=n1]&\phantom{aaaaaa}&[name=n2]\\
  \ncline[nodesep=0]{-]}{n}{n1}
  \Aput{\gamma}
  \ncline[nodesep=0]{-]}{n}{n2}
  \Aput{\beta}	
\endpsmatrix
&
\psmatrix
  &&&&&[name=n]\\\\\\\\\\\\
  &[name=n1]&\phantom{aaaaaa}&[name=n2]&\phantom{aa}&&\phantom{aa}&[name=n1']&\phantom{aaaaaa}&[name=n2']\\
  \ncline[nodesep=0]{-]}{n}{n1}
  \Aput{\alpha}
  \ncline[nodesep=0]{-]}{n}{n2}
  \Aput{\gamma}	
  \ncline[nodesep=0]{-]}{n}{n1'}
  \Aput{\gamma}
  \ncline[nodesep=0]{-]}{n}{n2'}
  \Aput{\beta}	
\endpsmatrix
\end{array}\]

By construction any path in either of the two dalographs is still a path in the new dalograph, so a non-maximal path is still a non-maximal path. By induction hypothesis, if there is a path that is $\prec$-maximal starting from the top node, then it induces $\beta$, but not $\alpha$. Hence, the new dalograph complies with the requirements.
\end{proof}

The A-transitive closure of a binary relation is its smallest A-transitive superrelation. It is inductively defined below.

\begin{defn}[A-transitive closure]
\[\begin{array}{c@{\hspace{.8cm}}c}
   \prooftree
    \alpha\prec\beta 
    \justifies
    \alpha\prec^{st}\beta 
  \endprooftree
&	
   \prooftree
    \alpha\prec^{st}\beta \qquad u\beta\prec^{st}\gamma 
    \justifies
    u\alpha\prec^{st}\gamma 
  \endprooftree
\end{array}\] 
\end{defn}

The following result about A-transitivity is a generalisation of the previous result about transitivity. 

\begin{lem}\label{lem:A-trans-step}
Let $\prec$ be a preference, and let $\alpha$ and $\beta$ be such that $\alpha\prec^{st}\beta$. There exists a dalograph $g$ with the following properties.

\begin{itemize}
\item The dalograph $g$ has the following shape (dashed lines represent and delimit the rest of the dalograph). 

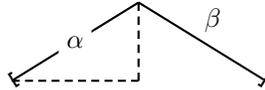
\[\psmatrix
  &&&[name=n]\\\\
  &[name=n1]&\phantom{aaaaaaaa}&[name=n1']&\phantom{aaaaaaaa}&[name=n2]\\
  \ncline[nodesep=0]{-]}{n}{n1}
  \mput*{\alpha}
  \ncline[nodesep=0,linestyle=dashed,dash=3pt 2pt]{-}{n}{n1'}
  \ncline[nodesep=0,linestyle=dashed,dash=3pt 2pt]{-}{n1}{n1'}
  \ncline[nodesep=0]{-]}{n}{n2}
  \Aput{\beta}	
\endpsmatrix\]

\item The path inducing $\beta$ is not branching after the top node.
\item Any equilibrium for $g$ involves the path inducing $\beta$.
\end{itemize}
\end{lem}

\begin{proof}
Proceed by rule induction on the definition of the A-transitive closure. First rule, the subproof is straightforward. Second rule, $u\alpha\prec^{st}\gamma$ comes from $u\beta\prec^{st}\gamma$ and $\alpha\prec^{st}\beta$. By induction hypothesis there exist a dalograph $g_1$ for $u\beta\prec^{st}\gamma$ and a dalograph $g_2$ for $\alpha\prec^{st}\beta$. Cut the path inducing $\beta$ away from $g_2$ (but the top node), and fuse two nodes as shown below.  

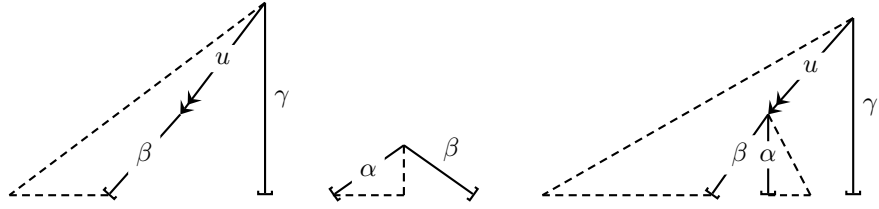
\[\begin{array}{l@{\hspace{.8cm}}l@{\hspace{.8cm}}r}
\psmatrix
  &&&&&&&&&[name=n]\\\\\\\\\\\\\\
  &&&&&&&[name=n1]&\phantom{aaaaa}&\\\\\\\\\\
  &[name=n2']&\phantom{aaaaaa}&[name=n2]&\phantom{aaa}&&&&&[name=n3]\\
  \ncline[nodesep=0]{->>}{n}{n1}
  \mput*{u}
  \ncline[nodesep=0]{-]}{n1}{n2}
  \mput*{\beta}
  \ncline[nodesep=0,linestyle=dashed,dash=3pt 2pt]{-}{n}{n2'}
  \ncline[nodesep=0,linestyle=dashed,dash=3pt 2pt]{-}{n2}{n2'}
  \ncline[nodesep=0]{-]}{n}{n3}
  \Aput{\gamma}
\endpsmatrix
&
\psmatrix
  &&&[name=n]\\\\\\
  &[name=n1]&\phantom{aaaa}&[name=n1']&\phantom{aaaa}&[name=n2]\\
  \ncline[nodesep=0]{-]}{n}{n1}
  \mput*{\alpha}
  \ncline[nodesep=0,linestyle=dashed,dash=3pt 2pt]{-}{n}{n1'}
  \ncline[nodesep=0,linestyle=dashed,dash=3pt 2pt]{-}{n1}{n1'}
  \ncline[nodesep=0]{-]}{n}{n2}
  \Aput{\beta}	
\endpsmatrix
&
\psmatrix
  &&&&&&&[name=n]\\\\\\\\\\\\
  &&&&&[name=n1]&\phantom{aaaaa}&\\\\\\\\\\
  &[name=n2']&\phantom{aaaaaaaaaaa}&[name=n2]&\phantom{aaa}&[name=n4]&[name=n5]&[name=n3]\\
  \ncline[nodesep=0,linestyle=dashed,dash=3pt 2pt]{-}{n}{n2'}
  \ncline[nodesep=0,linestyle=dashed,dash=3pt 2pt]{-}{n2}{n2'}
  \ncline[nodesep=0]{->>}{n}{n1}
  \mput*{u}
  \ncline[nodesep=0]{-]}{n1}{n2}
  \mput*{\beta}
  \ncline[nodesep=0]{-]}{n}{n3}
  \Aput{\gamma}
  \ncline[nodesep=0]{-]}{n1}{n4}
  \mput*{\alpha}
  \ncline[nodesep=0,linestyle=dashed,dash=3pt 2pt]{-}{n1}{n5}
  \ncline[nodesep=0,linestyle=dashed,dash=3pt 2pt]{-}{n4}{n5}
\endpsmatrix
\end{array}\]

By induction hypothesis, the node inducing $\gamma$ is not branching in $g_1$, so it it still not branching in the new dalograph $g$. In an equilibrium, the node just below $u$ must choose $\beta$, by induction hypothesis. So the top node must involve the path inducing $\gamma$, also by induction hypothesis. 
\end{proof}

The gen-E-prefix closure is a generalisation of the E-prefix closure.

\begin{defn}[gen-E-prefix closure]
\[\begin{array}{c@{\hspace{.8cm}}c}
   \prooftree
    \alpha\prec\beta 
    \justifies
    \alpha\prec^{sep}\beta 
  \endprooftree
  &
  \prooftree
    W\alpha\prec^{sep}W\beta
    \justifies
    \alpha\prec^{sep}\beta 
  \endprooftree
\end{array}\] 
\end{defn}

The following lemma is a step towards a generalisation of lemma~\ref{lem:epref-pres} about E-prefix closure. 

\begin{lem}\label{lem:sE-step}
Let $\prec$ be a preference. For any $\alpha\prec^{sep}\beta$ there exists a dalograph $g$ with the following properties.
\begin{itemize}
\item The dalograph below is a subgraph of the dalograph $g$. 

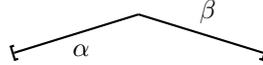
\[\psmatrix
  &&&[name=n]\\
  &[name=n1]&\phantom{aaaaaaaa}&&\phantom{aaaaaaaa}&[name=n2]\\
  \ncline[nodesep=0]{-]}{n}{n1}
  \Aput{\alpha}
  \ncline[nodesep=0]{-]}{n}{n2}
  \Aput{\beta}	
\endpsmatrix\]

\item Aside from the top node, the paths inducing $\alpha$ and $\beta$ are not branching.
\item Any equilibrium for $g$ involves the path inducing $\beta$.
\end{itemize}
\end{lem}

\begin{proof}
By rule induction on the definition of $\prec^{sep}$. First case, $\alpha\prec^{sep}\beta$ comes from $\alpha\prec\beta$. Straightforward. Second case, $\alpha\prec^{sep}\beta$ comes from $W\alpha\prec^{sep}W\beta$. By definition of $W\alpha\prec^{sep}W\beta$, for all $u$ in $W$ there exists $v$ in $W$ such that $u\alpha\prec v\beta$. So by induction hypothesis there exists a dalograph $g_{u,v}$ with the following properties.
\begin{itemize}
\item The dalograph below is a subgraph of the dalograph $g_{u,v}$. 

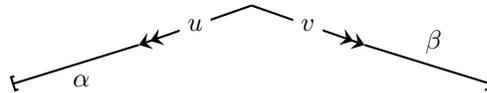
\[\psmatrix
  &&&&&[name=n]\\
  &&&[name=n1']&&\phantom{aaaaaaaaaaaaaa}&&[name=n2']\\
  &[name=n1]&\phantom{aaaaaaaa}&&&&&&\phantom{aaaaaaaa}&[name=n2]\\
  \ncline[nodesep=0]{->>}{n}{n1'}
  \mput*{u}
  \ncline[nodesep=0]{->>}{n}{n2'}
  \mput*{v}
  \ncline[nodesep=0]{-]}{n1'}{n1}
  \Aput{\alpha}
  \ncline[nodesep=0]{-]}{n2'}{n2}
  \Aput{\beta}	
\endpsmatrix\]

\item Aside from the top node, the paths inducing $u\alpha$ and $v\beta$ are not branching.
\item Any equilibrium for $g_{u,v}$ involves the path inducing $v\beta$.
\end{itemize}

Consider all these dalographs $g_{u,v}$ for $u$ in $W$. Fuse their top nodes into one node. Also fuse their nodes from where either $\alpha$ or $\beta$ starts into one single node. This yields a dalograph $g'$ with the following as a subgraph.

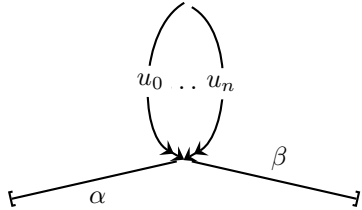
\[\psmatrix
  &&&[name=n0]\\\\\
  &&& \dots\\\\\
  &&&[name=n] \\
  &[name=n1]&\phantom{aaaaaaaaaa}&&\phantom{aaaaaaaaaa}&[name=n2]\\
  \ncarc[arcangleA=-60, arcangleB=-60,nodesep=0]{->>}{n0}{n}
  \mput*{u_0}
  \ncarc[arcangleA=60, arcangleB=60,nodesep=.1]{->>}{n0}{n}
  \mput*{u_n}
  \ncline[nodesep=.1]{-]}{n}{n1}
  \Aput{\alpha}
  \ncline[nodesep=.1]{-]}{n}{n2}
  \Aput{\beta}	
\endpsmatrix\]

Aside from the central node, the paths inducing $\alpha$ and $\beta$ are not branching. Each $u$ and $v$ are represented in the $u_i$, so any equilibrium for $g'$ involves the path inducing $\beta$.
\end{proof}

The following lemma states that under some conditions (similar to the conclusions of the lemmas above) equilibrium existence is preserved by superrelation.

\begin{lem}\label{lem:step-pres}
Let $\prec$ and $\prec'$ be two preferences. Assume that $\prec$ is included in $\prec'$ and that for any $\alpha\prec'\beta$, there exists a dalograph $g$ with the following properties.
\begin{itemize}
\item The dalograph below is a subgraph of the dalograph $g$. 
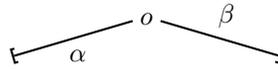
\[\psmatrix
  &&&[name=n]o\\
  &[name=n1]&\phantom{aaaaaaaa}&&\phantom{aaaaaaaa}&[name=n2]\\
  \ncline[nodesep=.1]{-]}{n}{n1}
  \Aput{\alpha}
  \ncline[nodesep=.1]{-]}{n}{n2}
  \Aput{\beta}	
\endpsmatrix\]
\item The path inducing $\beta$ is not branching after the top node $o$.
\item Any $\prec$-equilibrium for $g$ involves the path inducing $\beta$.
\end{itemize}
In this case, if all dalographs have $\prec$-equilibria, then all dalographs have $\prec'$-equilibria.
\end{lem}

\begin{proof}
Let $g$ be a dalograph. For each 3-uple $(o',\alpha,\beta)$ such that $\alpha\prec'\beta$ and $\alpha$ and $\beta$ are eligible at node $o'$ in $g$, do the following. By assumption, there is a dalograph $g_{o,\alpha,\beta}$ complying with the requirements below.
\begin{itemize}
\item The dalograph below is a subgraph of the dalograph $g_{o,\alpha,\beta}$. 
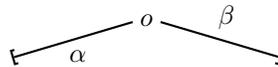
\[\psmatrix
  &&&[name=n]o\\
  &[name=n1]&\phantom{aaaaaaaa}&&\phantom{aaaaaaaa}&[name=n2]\\
  \ncline[nodesep=.1]{-]}{n}{n1}
  \Aput{\alpha}
  \ncline[nodesep=.1]{-]}{n}{n2}
  \Aput{\beta}	
\endpsmatrix\]
\item The path inducing $\beta$ is not branching after the top node $o$.
\item Any equilibrium for $g_{o,\alpha,\beta}$ involves the path inducing $\beta$.
\end{itemize}
Define $g'_{o,\alpha,\beta}$ by cutting away from $g_{o,\alpha,\beta}$ the path inducing $\beta$, but leaving the top node $o$. Fuse the node $o'$ from $g$ and the node $o$ from $g'_{o,\alpha,\beta}$. Let $g'$ be the dalograph built from $g$ after addition of all $g'_{o,\alpha,\beta}$. Let $s'$ be a $\prec$-equilibrium for $g'$. Let us consider again any 3-uple $(o',\alpha,\beta)$ such that $\alpha\prec'\beta$ and $\alpha$ and $\beta$ are eligible at node $o'$ in $g$. By construction at node $o'$, $s'$ does not induce any sequence that is eligible at the top node $o$ of $g'_{o,\alpha,\beta}$. More specifically, $s'$ does not induce $\alpha$ at node $o'$. Since this holds for any of the considered 3-uple, it means that $s'$ is also a $\prec'$-equilibrium. Removing the parts of $s'$ that corresponds to all the $g'_{o,\alpha,\beta}$ yields a $\prec'$-equilibrium for $g$.
\end{proof}

Thanks to the result above it is now possible to show that, like E-prefix closure, (A-) transitive closure, alt-subcontinuous closure, and gen-E-prefix closure preserve equilibrium existence.

\begin{lem}\label{lem:tutsc-pres}
If all dalographs have $\prec$-equilibria, then all dalographs have $\prec^{st}$-equilibria, $\prec^{sc}$-equilibria, $\prec^{sep}$-equilibria.
\end{lem}

\begin{proof}
By lemma~\ref{lem:step-pres} together with \ref{lem:A-trans-step} and \ref{lem:sE-step}.
\end{proof}

The combination closure of a binary relation is its smallest superrelation that is A-transitive and gen-E-prefix.

\begin{defn}[Combination closure]
\[\begin{array}{c@{\hspace{.8cm}}c@{\hspace{.8cm}}c}
   \prooftree
    \alpha\prec\beta 
    \justifies
    \alpha\prec^{c}\beta 
  \endprooftree
  &
  \prooftree
    \alpha\prec^{c}\beta \qquad u\beta\prec^{c}\gamma 
    \justifies
    u\alpha\prec^{c}\gamma 
  \endprooftree
& 
\prooftree
    W\alpha\prec^{c}W\beta
    \justifies
    \alpha\prec^{c}\beta 
  \endprooftree
\end{array}\] 

\end{defn}

The combination closure preserves equilibrium existence, as stated below.

\begin{thm}\label{thm:fc-pres}
If all dalographs have $\prec$-equilibria, all dalographs have $\prec^{c}$-equilibria.
\end{thm}

\begin{proof}
Let $\prec$ be a preference that guarantees existence of equilibrium. Let $g$ be a dalograph and let $S$ be the finite set of all pairs of sequences that are eligible in $g$. The A-transitive closure and gen-E-prefix closure are simple closures, and they preserve equilibrium existence by lemmas~\ref{lem:tutsc-pres}. Moreover, the combination closure is their union, so by lemma~\ref{lem:union-pres}, the restriction to $S$ of the full closure of $\prec$ also guarantees existence of equilibrium. Since $\prec^{c}|_S$-equilibrium is also a $\prec^{c}$-equilibrium by lemma~\ref{lem:coincide-eq}, this allows concluding. 
\end{proof}

\section{Sufficient Condition and Necessary Condition}\label{sect:sc-nc}

After a synthesis of and a discussion about the results obtained so far, this section gives a non-trivial example of a relation that does not meet the requirement of the necessary condition for equilibrium existence. Finally, a further result on network routing application is given.
 
\subsection{Synthesis}

Firstly, this subsection gathers the main results of this chapter that concern existence of equilibrium. Secondly, It points out that if the preference is a total order, then the sufficient condition and the necessary conditions coincide. Finally, it shows that the necessary condition is not sufficient in general.

The following theorem presents the sufficient condition and the necessary condition for every dalograph to have an equilibrium. The sufficient condition involving the notion of strict weak order is written with few words. However it is difficult to compare it with the necessary condition. Therefore, the sufficient condition is rewritten in a way that enables comparison.

\begin{thm}\label{thm:syn}
\[\begin{array}{c}
\mbox{The preference $\prec$ is included in some $\prec'$.}\\
\mbox{The preference $\prec'$ is an E-prefix and subcontinuous strict weak order.}\\
\Updownarrow\\
\mbox{The preference $\prec$ is included in some $\prec'$.}\\
\mbox{The preference $\prec'$ is E-prefix, subcontinuous, transitive, and irreflexive.}\\
\mbox{The non comparability relation $\sharp'$ is transitive.}\\
\Downarrow\\
\mbox{Every dalograph has a $\prec$-equilibrium.}\\
\Downarrow\\
\mbox{The preference $\prec$ is included in some $\prec'$.}\\
\mbox{The preference $\prec'$ is (gen-) E-prefix, (A-) transitive, and irreflexive.}\\
\end{array}\]
\end{thm}

\begin{proof}
The topmost two propositions are equivalent by lemma~\ref{lem:kn-teq-ic}, and they imply the third proposition by lemma~\ref{thm:ESkr-eq}. For the last implication, assume that all dalographs have $\prec$-equilibria. By theorem~\ref{thm:fc-pres}, all dalographs have $\prec^{c}$-equilibria. So $\prec^{c}$ is irreflexive, otherwise the reflexive witness alone allows building a game without $\prec^{c}$-equilibrium. In addition, $\prec^{c}$ is E-prefix and A-transitive, and $\prec$ is included in $\prec^{c}$ by construction. 
\end{proof}

When the preference is a strict total order, the following corollary proves a necessary and sufficient condition for all dalographs to have equilibria.  

\begin{cor}\label{cor:to}
Let a preference be a strict total order. All dalographs have equilibria \textit{iff} the preference is E-prefix and subcontinuous.
\end{cor}

\begin{proof}
Left-to-right implication: by theorem~\ref{thm:syn}, the strict total order is gen-E-prefix and A-transitive, so it is E-prefix and transitive. If $\prec$ is not subcontinuous, $u^\omega\prec\alpha\prec u\alpha$ for some $u$ and $\alpha$. So $u$ is non-empty and $\alpha\prec uu\alpha$, and the following dalograph has no equilibrium.

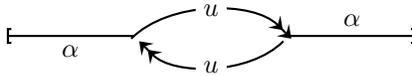
\[\psmatrix\\
  &[name=n2]&\phantom{aaaaaaaa}&[name=n1]&\phantom{aaaaaaaaaa}&[name=n]&\phantom{aaaaaaaa}&[name=n3]\\\\
  \ncline[nodesep=0]{-]}{n1}{n2}
  \Aput{\alpha}
  \ncarc[arcangleA=45, arcangleB=45]{->>}{n1}{n}
  \mput*{u}
  \ncarc[arcangleA=45, arcangleB=45]{->>}{n}{n1}
   \mput*{u}
  \ncline[nodesep=0]{-]}{n}{n3}
  \Aput{\alpha}
\endpsmatrix\]

The right-to-left implication follows directly theorem~\ref{thm:syn} because a strict total order is a strict weak order.
\end{proof}

There is a direct proof of this corollary, some parts of whose are much simpler than the proof of theorem~\ref{thm:syn}. For instance, if a total order is E-prefix, then its negation is also E-prefix. This ensures that any maximal path is also semi-hereditary maximal. Therefore the definition of the seeking-forward function is not needed. The necessary condition is much simpler too. Indeed, if $\prec$ is not E-prefix, we have $u\alpha\prec u\beta$ and $\alpha\not\prec\beta$ for some $u$, $v$, $\alpha$ and $\beta$. By the first assumption we have $\alpha\neq\beta$, so $\beta\prec\alpha$ by total ordering. Therefore the following dalograph has no equilibrium.

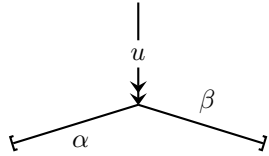
\[\psmatrix\\
  &&&[name=n0]\\\\\\
  &&&[name=n]\\
  &[name=n1]&\phantom{aaaaaaaa}&&\phantom{aaaaaaaa}&[name=n2]\\
  \ncline[nodesep=.1]{->>}{n0}{n}
  \mput*{u}
  \ncline[nodesep=0]{-]}{n}{n1}
  \Aput{\alpha}
  \ncline[nodesep=0]{-]}{n}{n2}
  \Aput{\beta}	
\endpsmatrix\]

In general, the necessary condition is not a sufficient condition, as shown by the following two examples. First, let $\prec$ be defined as followed.

\[\begin{array}{r@{\hspace{.2cm}\prec\hspace{.2cm}}l@{\hspace{.8cm}}r@{\hspace{.2cm}\prec\hspace{.2cm}}l}
u_1y_1\beta_2  &  v_1x_1\alpha_1  &  u_2y_2\beta_1  &  v_2x_2\alpha_2\\
v_1x_2\alpha_2  &  u_1\alpha_1  &  v_2x_1\alpha_1  &  u_2\alpha_2\\
v_1x_1y_1\beta_2  &  u_1\alpha_1  &  v_2x_2y_2\beta_1  &  u_2\alpha_2 \\
v_1x_2y_2\beta_2  &  u_1\alpha_1  &  v_2x_1y_1\beta_1  &  u_2\alpha_2
\end{array}\]

The preference $\prec$ complies with the necessary condition but the dalograph below has no $\prec$-equilibrium. Indeed, the node $o_1$ "wants" to follow a path leading to $\alpha_1$ or $\beta_1$, while the node $o_2$ "wants" to follow a path leading to $\alpha_2$ or $\beta_2$.

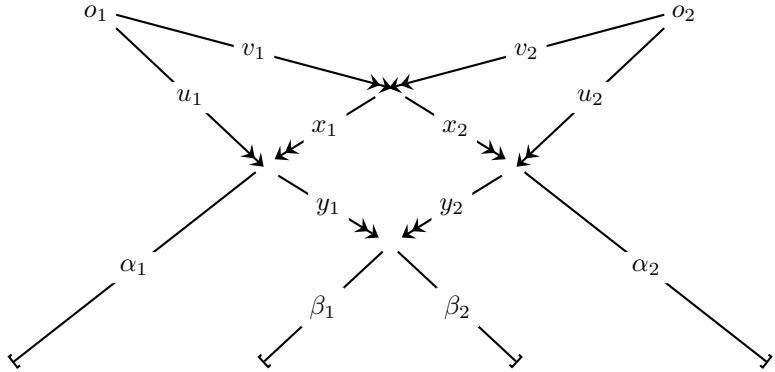
\[\psmatrix\\
  &&&[name=n1]o_1&\phantom{aaaaaaaaaa}&&\phantom{aaaaaaaa}&&\phantom{aaaaaaaa}&&\phantom{aaaaaaaaaa}&[name=n2]o_2\\\\
  &&&&&&&[name=n3] \\\\
  &&&&&[name=n4]&&&&[name=n5]\\\\
  &&&&&&&[name=n6] \\\\\\
  &[name=n7]&\phantom{aaaa}&&&[name=n8]&&&&[name=n9]&&&\phantom{aaaa}&[name=n10]\\
  \ncline[nodesep=.1]{->>}{n1}{n3}
  \mput*{v_1}
  \ncline[nodesep=.1]{->>}{n1}{n4}
  \mput*{u_1}
  \ncline[nodesep=.1]{->>}{n2}{n3}
  \mput*{v_2}
  \ncline[nodesep=.1]{->>}{n2}{n5}
  \mput*{u_2}
  \ncline[nodesep=.2]{->>}{n3}{n4}
  \mput*{x_1}
  \ncline[nodesep=.2]{->>}{n3}{n5}
  \mput*{x_2}
  \ncline[nodesep=.2]{->>}{n4}{n6}
  \mput*{y_1}
  \ncline[nodesep=.2]{->>}{n5}{n6}
  \mput*{y_2}
  \ncline[nodesep=.1]{-]}{n4}{n7}
  \mput*{\alpha_1}
  \ncline[nodesep=.1]{-]}{n6}{n8}
  \mput*{\beta_1}
  \ncline[nodesep=.1]{-]}{n6}{n9}
  \mput*{\beta_2}
  \ncline[nodesep=.1]{-]}{n5}{n10}
  \mput*{\alpha_2}
\endpsmatrix\]

Second example, let $\prec$ be defined as followed. 
\[\begin{array}{r@{\hspace{.2cm}\prec\hspace{.2cm}}l@{\hspace{.8cm}}r@{\hspace{.2cm}\prec\hspace{.2cm}}l@{\hspace{.8cm}}r@{\hspace{.2cm}\prec\hspace{.2cm}}l}
 \alpha_1 & y_1x_2\alpha_2  &  \alpha_2 & y_2x_3\alpha_3 &  \alpha_3 & y_3x_1\alpha_1\\
 (x_1y_1)^\omega & x_3\alpha_3  &  (x_2y_2)^\omega & x_1\alpha_1  &  (x_3y_3)^\omega & x_2\alpha_2 
\end{array}\]

The preference $\prec$ complies with the necessary condition but the dalograph below has no $\prec$-equilibrium. Indeed, the situation looks like it is in the jurisdiction of alt-subcontinuity, but it is not.

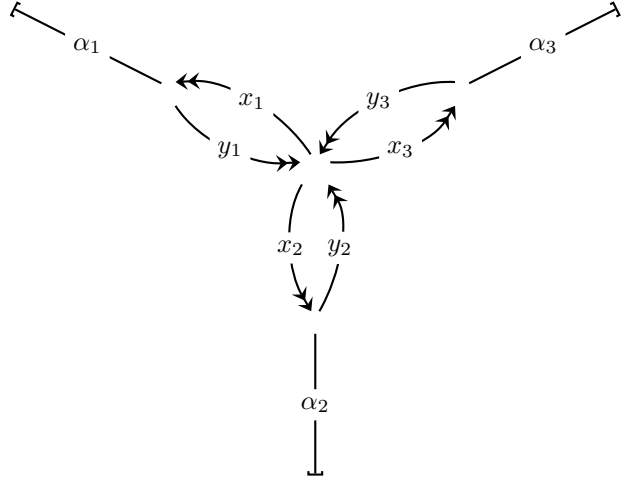
\[\psmatrix\\
 &[name=n1]&\phantom{aaaaaaaaaa}&&&&&&\phantom{aaaaaaaaaa}&[name=n2]\\\\
 &&&[name=n3]&\phantom{aaaaaaaaaa}&&\phantom{aaaaaaaaaa}&[name=n4]\\\\
 &&&&&[name=n5]\\\\\\\\
 &&&&&[name=n6]\\\\\\\\
 &&&&&[name=n7]\\
 \ncline[nodesep=.1]{-]}{n3}{n1}
 \mput*{\alpha_1}
 \ncline[nodesep=.1]{-]}{n4}{n2}
 \mput*{\alpha_3}
 \ncline[nodesep=.1]{-]}{n6}{n7}
 \mput*{\alpha_2}
 \ncarc[arcangleA=-30, arcangleB=-30,nodesep=.2]{->>}{n5}{n3}
 \mput*{x_1}
 \ncarc[arcangleA=-30, arcangleB=-30,nodesep=.2]{->>}{n5}{n4}
 \mput*{x_3}
 \ncarc[arcangleA=-30, arcangleB=-30,nodesep=.2]{->>}{n5}{n6}
 \mput*{x_2}
 \ncarc[arcangleA=-30, arcangleB=-30,nodesep=.2]{->>}{n3}{n5}
 \mput*{y_1}
 \ncarc[arcangleA=-30, arcangleB=-30,nodesep=.2]{->>}{n4}{n5}
 \mput*{y_3}
 \ncarc[arcangleA=-30, arcangleB=-30,nodesep=.2]{->>}{n6}{n5}
 \mput*{y_2}
\endpsmatrix\]

It is possible to design closures that rule out the above "annoying" situations. For instance, the closure suggested below (by mutual induction) may take care of the triskele example (and of any related example with n branches). However, this kind of incremental procedure is very likely to leave out some more complex examples.

\[\begin{array}{c@{\hspace{.8cm}}c@{\hspace{.8cm}}c}
   \prooftree
    \alpha\prec\beta 
    \justifies
    \alpha\prec^{l}\beta 
  \endprooftree
  &
  \prooftree
    \alpha\prec^{l} v\beta \qquad (uv)^\omega\prec^{l}\gamma 
    \justifies
    u\alpha\prec^{ll}\beta,\gamma 
  \endprooftree
  &
  \prooftree
    \alpha\prec^{ll}\beta 
    \justifies
    \alpha\prec^{l}\beta
\endprooftree
  \end{array}\]
\[\begin{array}{c@{\hspace{.8cm}}c@{\hspace{.8cm}}c@{\hspace{.8cm}}c}
 \prooftree
    A\prec^{ll}B \qquad B\prec^{ll} C
    \justifies
    A\prec^{ll}C 
  \endprooftree   
&
\prooftree
    A\prec^{l}B 
    \justifies
    A,C\prec^{l} B,C 
  \endprooftree
&
   \prooftree
    A,C\prec^{ll}B,C 
    \justifies
    A\prec^{l}B
  \endprooftree 
  \end{array}\]
\medskip

\subsection{Example}

The max-min limit-set order is defined in subsection~\ref{subsect:ex-eq-ex}, whereas the max-min set order is defined below. It does not only consider the limit set of the sequence, but every element occurring in the sequence.

\begin{defn}[Max-min set order] Let $(E,\prec)$ be a set equipped with a strict weak order. For $\alpha$ an infinite sequence over $E$, let $S_\alpha$ be the set of all elements occurring in $\alpha$.
\[\alpha\prec^{Mms}\beta\quad\eqdef\quad S_\alpha\prec^{Mm}S_\beta\]
\end{defn}

This order cannot guarantee existence of global equilibrium, as stated below.

\begin{lem}
There exists $(E,\prec)$ a set equipped with a strict weak order, such that there exists a dalograph that is labelled with elements in $E$, and that has no global equilibrium with respect to the max-min set order.
\end{lem}

\begin{proof}
Along the usual order over the figures $0$, $1$ and $2$, we have $2(02)^\omega\prec^{Mms}21^\omega$, since $\{0,2\}\prec^{Mm}\{1,2\}$. However, when removing the first $2$ of these two sequences, we have $1^\omega\prec^{Mms}(02)^\omega$ since $\{1\}\prec^{Mm}\{0,2\}$. Therefore any E-prefix and transitive relation including $\prec^{Mms}$ is not irreflexive. Conclude by theorem~\ref{thm:syn}.
\end{proof}

\subsection{Application to Network Routing}

In the total order case, the necessary and sufficient condition for equilibrium in dalographs yields a necessary and sufficient condition for routing equilibrium in routing problems. The necessary condition implication invokes constructive arguments that are similar to the ones used for the necessary condition in dalographs. However, the proof is simple enough so that just doing it is more efficient than applying a previous result.

\begin{thm}
Assume a routing policy $\prec^r$ that is a total order. Then every routing problem has a routing equilibrium \textit{iff} the policy is E-prefix and $uv\prec^rv$ for all $v$ and non-empty $u$.
\end{thm}

\begin{proof}
Left-to-right: by contraposition, assume that either $\prec^r$ is not E-prefix or there exists $u$ and $v$ such that  $v\prec^ruv$. First case, $\prec^r$ is not E-prefix. So there exists $u$, $v$ and $w$ such that $uv\prec^ruw$ and $w\prec^rv$. So the following routing problem has no routing equilibrium.

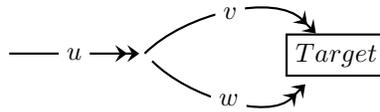
\[\psmatrix
  &[name=n] &\phantom{aaaaaaaaa}&[name=n1] &\phantom{aaaaaaaaa}&[name=n2]\psframebox{Target}\\\\
  \ncline{->>}{n}{n1}
  \mput*{u}
  \ncarc[arcangleA=45, arcangleB=45,nodesep=0]{->>}{n1}{n2}
  \mput*{v}
  \ncarc[arcangleA=-45, arcangleB=-45,nodesep=.1]{->>}{n1}{n2}
  \mput*{w}	
\endpsmatrix\]

Second case, there exists $u$ and $v$ such that $v\prec^ruv$. The following routing problem has no routing equilibrium.

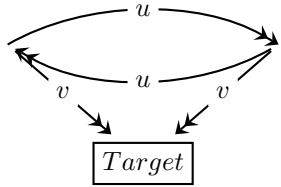
\[\psmatrix\\\\
  &[name=n] &\phantom{aaaaa}&&\phantom{aaaaa}& [name=n1]\\\\\\
  &&&[name=n2]\psframebox{Target}\\
  \ncarc[arcangleA=30, arcangleB=30,nodesep=0]{->>}{n}{n1}
  \mput*{u}
  \ncarc[arcangleA=30, arcangleB=30,nodesep=.1]{->>}{n1}{n}
  \mput*{u}
  \ncline[nodesep=.1]{->>}{n}{n2}
  \mput*{v}
  \ncline[nodesep=.1]{->>}{n1}{n2}
  \mput*{v}	
\endpsmatrix\]
The right-to-left implication follows lemma~\ref{lem:routing-eq-impl}.
\end{proof}

\section{Conclusion}

Consider a collection of labels and a binary relation, called preference, over ultimately periodic sequences over these labels. This chapter shows that if the preference is an E-prefix and subcontinuous strict weak order, then all dalographs labelled with those labels have equilibria with respect to this preference. This sufficient condition is proved by a recursively-defined seeking-forward function followed by a proof by induction on the number of arcs in a dalograph. A necessary condition is also proved thanks to the notion of simple closure and the design of some simple closures, the union of which preserves equilibrium existence. Some examples show that the necessary condition is not sufficient in general. However, a few examples show the usefulness of both the sufficient and the necessary conditions. A detailed study shows that the neccessary condition plus the subcontinuity plus the transitivity of the incomparability relation implies the sufficient condition. Because of this, the two conditions coincide when the preference is a strict total order, which could also be found by a direct proof. However for now, there is no obvious hint saying whether or not the sufficient condition is also necessary. 

This chapter applies its theoretical results to a network routing problem: first, the above-mentioned sufficient condition yields a sufficient condition on routing policy for routing equilibrium existence in a simple routing problem. Second, the above-mentioned necessary and sufficient condition of the total order case also yields a necessary and sufficient condition on a total order routing policy for routing equilibrium existence in a simple routing problem. 

This chapter is also useful to one other respect: many systems that are different from dalographs also require a notion of preference. In a few of these systems, preferences may be thought as total orders without a serious loss of generality: in these systems, any preference that guarantees equilibrium existence is included in some total order also guaranteeing equilibrium existence, and equilibrium existence is preserved by subrelation. In such a setting, considering only total orders somehow accounts for all binary relations. However in the case of dalographs, there might exist a preference guaranteeing equilibrium existence, such that any linear extension of the preference does not guarantee equilibrium existence. In this case, assuming total ordering of the preference would yield a (non-recoverable) loss of generality. The following example is a candidate for such a preference. Consider the ultimately periodic sequences over $\{a,b,c,d\}$. An A-transitive preference $\prec$ over these sequences is defined below.

\[\begin{array}{r@{\hspace{.2cm}\prec\hspace{.2cm}}c@{\hspace{.2cm}\prec\hspace{.2cm}}l}
  a^\omega & cb^\omega & da^\omega \\
 db^\omega & ca^\omega & b^\omega
\end{array}\]
 
The preference $\prec$ defined above is (A-) transitive and (gen-) E-prefix. In addition, it is not included in any transitive and E-prefix total order. Indeed, let $<$ be such a total order. If $a^\omega<b^\omega$ then $da^\omega<db^\omega$ by E-prefix and total ordering, so $cb^\omega<ca^\omega$ by transitivity, so $b^\omega<a^\omega$ by E-prefix, contradiction. If $b^\omega<a^\omega$ then $ca^\omega<cb^\omega$ by transitivity, so $a^\omega<b^\omega$ by E-prefix, contradiction. So, the key question is whether or not the preference $\prec$ guarantees equilibrium existence for all dalographs (this is not proved in this chapter).

\bibliographystyle{plain}
\bibliography{graph_optim_report}

\end{document}